\documentclass{article}
\usepackage{times}

\usepackage{soul}
\usepackage{url}
\usepackage[hidelinks]{hyperref}
\usepackage[utf8]{inputenc}
\usepackage[small]{caption}
\usepackage{graphicx}
\usepackage{amsmath}
\usepackage{booktabs}
\usepackage{style_arxiv}
\title{Robust Solutions for Multi-Defender Stackelberg Security  Games\thanks{This research has been partly supported
by the Israel Science Foundation under grant 1958/20 and the EU Project TAILOR
under grant 992215.}}

\makeatletter
    \let\@internalcite\cite
    \def\cite{\def\citeauthoryear##1##2{##1, ##2}\@internalcite}
    \def\shortcite{\def\citeauthoryear##1{##2}\@internalcite}
    \def\@biblabel#1{\def\citeauthoryear##1##2{##1, ##2}[#1]\hfill}
\makeatother

\begin{document}

\title{Robust Solutions for Multi-Defender Stackelberg Security  Games\thanks{This research has been partly supported
by the Israel Science Foundation under grant 1958/20 and the EU Project TAILOR
under grant 952215.}}

\author{Dolev Mutzari Yonatan Aumann Sarit Kraus\\
Department of Computer Science, Bar Ilan University, Ramat Gan, Israel \\
dolevmu@gmail.com, aumann@cs.biu.ac.il, sarit@cs.biu.ac.il}
\date{}



\maketitle


\begin{abstract}
Multi-defender Stackelberg Security Games (MSSG) have recently gained increasing attention in the literature. 
However, the solutions offered to date are highly sensitive, wherein even small perturbations in the attacker's utility or slight uncertainties thereof can dramatically change the defenders' resulting payoffs and alter the equilibrium. 
In this paper, we introduce a robust model for MSSGs, which admits solutions that are resistant to small perturbations or uncertainties in the game's parameters. 
First, we formally define the notion of robustness, as well as the robust MSSG model. Then, for the non-cooperative setting, we prove the existence of a robust approximate equilibrium in any such game, and provide an efficient construction thereof. For the cooperative setting, we show that any such game admits a robust approximate $\alpha$-core, provide an efficient construction thereof, and prove that stronger types of the core may be empty.
Interestingly, the robust solutions can substantially increase the defenders' utilities over those of the non-robust ones.

\end{abstract}

\section{Introduction}



Stackelberg Security Games (SSGs) have attracted much attention in multi-agent community~\cite{paruchuri2008efficient,tambe2011security,nguyen2013analyzing,an2016stackelberg}. 
They were applied to a variety of security problems, including the allocation of security resources at the Los Angeles International Airport~\cite{pita2008deployed}, and protecting biodiversity in conservation areas~\cite{basak2016abstraction} (see~\cite{sinha2018stackelberg} for an overview).
The original SSG model postulates a single defender facing a single attacker. In many applications, however, different targets may be valuable --- to varying extents --- to multiple disparate parties, each of which is interested in defending its targets of interest. Accordingly, the model of Multi-defender Stackelberg Security Games (MSSGs), wherein multiple defenders (leaders) protect a set of targets against a strategic attacker (follower) who is their common enemy, has recently gained increasing attention. Solution concepts for this game were developed over time, starting from $\zeta$-Nash Equilibrium (NE)~\cite{gan2018stackelberg}, coordination mechanism~\cite{gan2020mechanism}, coalition formation~\cite{mutzari2021coalition} and correlated equilibrium (CE) via negotiation~\cite{castiglioni2021committing}.

However, for reasons we will detail shortly, to date, the solutions offered for MSSGs are highly sensitive, wherein small perturbations in the attacker's utility, or slight uncertainties thereof, can dramatically alter the defenders' resulting payoffs and alter the equilibrium. 

\paragraph{Tie-Breaking and Discontinuity.}
The key, and somewhat subtle, reason for the high sensitivity of the previous solutions has to do with the issue of tie-breaking. 
Typically, in SSGs, at equilibrium, the attacker has several equally attractive targets. Indeed, otherwise the defender(s) could shift resources from the overly protected targets to increase the protection of the attacker's chosen target. So, as ties are ubiquitous, the attacker's tie-breaking behaviour plays a crucial role in the analysis. Different works have considered different tie-breaking policies. In the case of a single defender, it is commonly assumed that the attacker breaks ties in favor of the defender, the reason being that the defender can essentially enforce his choice by shifting an arbitrarily small amount of resources away from his desirable target, absorbing only an arbitrarily small utility loss.

In the case of multiple heterogeneous defenders, however, optimistic tie-breaking is not well-defined (and not justified), as the best target for each defender may be different. Hence, \cite{gan2018stackelberg} use \emph{pessimistic tie-breaking}, wherein each defender acts as if (among its choices) the attacker will attack the target that is \emph{worst} for the defender. So, different defenders act as if the attacker, deterministically, attacks a different target. This tie-breaking rule, however, is not only overly pessimistic and unrealistic (as the attacker cannot simultaneously attack multiple targets), but also results in a sharp non-continuity in the defenders' payoffs. This is because the pessimistic tie-breaking occurs only at the point of \emph{exactly identical} utilities (of the attacker) for the disparate targets. If the attacker's utility is slightly perturbed (or if there is even a slight uncertainty with regards to the attacker's true utilities), then the pessimistic tie-breaking is no longer relevant and the defenders' payoff may change dramatically. 
This discontinuity in payoff, in turn, shapes the structure of the solution concepts, which also tend to exhibit a discontinuity at the exact equilibrium point. Indeed, this sensitivity, and discontinuity, is the core reason why, in the existing MSSG models, exact-NE need not exist (see Example~\ref{example:no-wse}).

In practice, however, such sharp discontinuity is often hard to justify. Indeed, while it is standard in game theory to assume that players' utilities are common knowledge, it is unrealistic to assume that these utilities are known accurately to any level of precision. Evidently, the defenders must somehow \emph{infer} the attacker's utilities using some combination of data and reasoning, but this data (and reasoning) may be incomplete, imprecise, or noisy. Additionally, the attacker itself may have bounded computational resources, or be subject to other forms of noise, and thus not operate exactly as dictated by infinite precision calculations. Finally, the attacker's strategy is also determined by its knowledge/belief of the defender's coverage strategies, which the attacker needs to somehow infer from possibly incomplete and noisy data.  

\paragraph{Robustness in MSSGs.} 
Accordingly, as detailed in the related work, in the \emph{single} defender setting, a whole line of research is devoted to addressing uncertainties (see \cite{nguyen2014stop}). Most of these papers tackle uncertainty by applying robust optimization techniques
(\cite{Pita2009HumanUncertainties,Jiang20131MonotonicMaximin,qian2015robust,nguyen2014stop}).
Some works consider a Bayesian approach \cite{kiekintveld2011approximation,Yin2012UnifiedDiscreteContinuous,yang2012computing,amin2016gradient}.
We believe, however, that this avenue is ill-suited for the multi-defender setting, with its tie-breaking subtlety. Robust optimization takes a \emph{worst-case} approach: if some parameters are only known to lay within some set, then one assumes --- pessimistically --- that these parameters obtain values that minimize the objective function. In the multi-defender setting, however, there is no \emph{one} worst case; the worst-case for one defender may well be good for another. So, worst-case robustness is not well defined in this setting. We note that one could possibly propose a multi-player worst case analysis wherein each defender acts as if the parameters are worst for itself.  But, coupled with pessimistic tie-breaking, this would result in an unreasonably, doubly pessimistic and unrealistic perspective, wherein different defenders postulate attackers with different parameters, attacking different targets; as if they inhabit parallel universes. 
The multi-defender setting therefore calls for a different approach. 

\paragraph{Our Contribution.} In this paper, we offer a robust model for the analysis of MSSGs.
First, we formally define the notion of a \emph{robust solution}, which formalizes the notion that the game solutions (e.g. Nash equilibrium, core) remain valid even after small perturbations or uncertainties in the game's parameters. We then introduce a formal model wherein small perturbations in the attacker's utility result in only small changes in the attacker's expected behavior, and hence also in that of the defenders. Essentially, we model the attacker's behavior as being probabilistic, with a continuous distribution concentrated around the behavior dictated by the presumed (and possibly inaccurate) utility. Importantly, we do not suppose any specific form for this distribution, only that it is concentrated as stated. Thus, this one model captures and unifies many possible scenarios and sources of noise and uncertainty.

Once we have formally defined the notions of robustness and the robust MSSG model, we show that the robust model indeed offers robust solutions. 
For the non-cooperative setting, we provide an efficient algorithm for constructing a robust approximate NE. For the cooperative setting, we consider the core of the game, for different variants of the core ($\alpha, \gamma$). For the $\alpha$ version, we prove that the robust  approximate core is always non-empty, and give an efficient algorithm for finding solutions therein. Importantly, since the utility of the defenders is no longer determined by their pessimistic beliefs, the resulting core solutions allow greater utility for the defenders than in previous models. Finally, we show that the $\gamma$-core may be empty. 

We note that an additional benefit of the model, besides robustness, is that it renders moot the entire issue of tie-breaking --- optimistic or pessimistic. With a continuous probability distribution, there is zero probability for ties.

\paragraph{Selected Related Work.} Robust analysis of the single-defender case has attracted considerable attention. Many of these employ robust optimization techniques to address uncertainties, while others model uncertainties using Bayesian models.
Robust optimization is employed by: \cite{Kiekintveld2013Interval} in a setting where the attacker's utilities are known to lay in some interval, but the exact value is not known; and by
\cite{Pita2009HumanUncertainties} is a settings wherein the attacker may exhibit bounded rationality; by \cite{Jiang20131MonotonicMaximin} in a setting wherein the attacker's type is only known to be monotone.  
\cite{qian2015robust} consider a worst-case approach for studying risk-averse attackers, for which the level of risk-adverseness is unknown.
Finally, \cite{nguyen2014stop} offer a unified framework and methods for simultaneously addressing multiple types of uncertainties in single defender SSGs using robust optimization. 
As detailed in the introduction, we argue that the worst-case approach of robust optimization is ill-suited for the multi-defender setting, wherein there is no one worst-case for all defenders. 

Other works, still in the single defender setting, employ a Bayesian approach to address uncertainties (as we do for the multi-defender setting).
A general analysis of a single defender Bayesian SSG was introduced in~\cite{paruchuri2008playing}.
\cite{kiekintveld2011approximation} study Bayesian SSGs with a continuous payoff distribution for the attacker. 
They demonstrate scenarios where there are too many possible attacker types to consider, and known solutions for a finite number of attacker types don't scale well.
\cite{Yin2012UnifiedDiscreteContinuous} provide a unified Bayesian approach to handling both discrete and continuous uncertainty in attacker types. \cite{yang2012computing} model human adversaries with bounded rationality as quantal response adversaries%
, which with some probability do not choose the best response. 
They specifically assume that the attacker's strategy is determined by a scaled \emph{soft-max} function of the expected utilities of the attacker on each target, and utilize the specific structure of the \emph{soft-max} function to obtain a solution for the setting (see also \cite{amin2016gradient}, who consider general SGs). 

Our work differs from all of the above in that it considers the multi-defender case. The Bayesian modeling we introduce is also different.  On the one hand, unlike the unrestricted, general Bayesian models (e.g. ~\cite{paruchuri2008playing,kiekintveld2011approximation,Yin2012UnifiedDiscreteContinuous}), 
~\cite{paruchuri2008playing,Yin2012UnifiedDiscreteContinuous}), 
we assume a concentrated distribution (Definition \ref{def:ed-ABF}). This, we believe, adequately models most frequent sources of uncertainty, including noisy and imprecise information, bounded computation power, and most incarnations of bounded rationality. On the other hand, we do not suppose any specific form for this concentrated distribution (e.g. \emph{soft-max}), as such an assumption would fail to model many real-world uncertainties.

\section{Defining Robustness}
The main objective of this work is to develop a robust MSSG framework. 
We now formally define the notion of robustness. We deliberately do so in the most general terms, not confining the definition to the specific game, or solution concepts that we consider. We later show how the model we offer indeed exhibits robustness as defined here.  

\paragraph{Solution Concepts.}
Consider the Nash equilibrium solution concept. Technically, given a game $G$, there is a set $\mbox{NE}(G)$ of strategy profiles for which the NE property holds.  Similarly, the core of $G$ is a set of coalition structures $\mbox{\it core}(G)$, for which the core properties hold. Thus, in the most general sense, a \emph{solution concept} is a function $\mathcal{X}$ from the set of games (of some class) to some space $S$, which maps each game $G$ (of the class) to the corresponding structures for which the specific property of the solution concept holds. 

\paragraph{Nearness.}
Conceptually, robustness of a solution $\xx$ states that $\xx$ remains a solution even under ``small'' perturbations in game parameters. This requires a notion of \emph{distance} amongst games. Given a distance function $d(\cdot,\cdot)$ over game pairs, say that games $G,\hat{G}$ are \emph{$\eta$-near} if $d(G,\hat{G})\leq \eta$ (later, $d$ will be instantiated based on the MSSG's specific parameters).
\begin{Definition}[\bf Robust Solution]
\label{def:robust-solution}
Let $\mathcal{X}$ be a solution concept, $G$ a game, and $x\in \mathcal{X}(G)$.  We say that $x$ is an \emph{$\eta$-robust $\mathcal{X}$ of $G$}, if $x\in \mathcal{X}(\hat{\GG})$ for any $\hat{G}$ that is $\eta$-near to $G$.
\end{Definition}
Thus, for example, a strategy profile $\xx$ is said to be an $\eta$-robust $\epsilon$-NE of $G$ if it is an $\epsilon$-NE of any $\hat{G}$ that is $\eta$-near to $G$ ($G$ itself included).


\section{The Robust MSSG Model}
\paragraph{The Standard model.}
In an MSSG, there is a set $\calN=\{1,\dots,n\}$ of {\em defenders}, a set $\calT=\{t_1,\ldots,t_m\}$ of {\em targets} that the defenders wish to protect, and an {attacker}, who seeks to attack the targets. 
Each defender $i\in \calN$ has $k_i \in \mathbb{N}$ {\em security resources}, each of which can be allocated to \emph{protect} a target. 
The attacker chooses one target to attack. The attack is \emph{successful} if the target is unprotected by any security resource.   A successful (res. unsuccessful) attack at $t$ yields utility $r^a(t)$ (res. $p^a(t)$) to the attacker and $p^d_i(t)$ (res. $r^d_i(t)$) to defender $i$, for all $i$ ($r^d_i(t)> p^d_i(t)$, $r^a(t)> p^a(t)$).

An MSSG is thus a 5-tuple $G=(\calN,\calT,\mathcal{K}, \mathcal{R},\calP)$, where $\mathcal{K}=(k_1,\ldots,k_n)$ are the numbers of security resources of the defenders, $\mathcal{R}$ is the sequence  of rewards, $\mathcal{R} =(r^a(t_1),\ldots,r^a(t_m),r^d_1(t_1),\ldots,r^d_n(t_m))$, and similarly $\calP$ is the sequence of penalties.

The defenders' allocation of security resources to the targets may be randomized. Specifically, each defender $i$ chooses a \emph{coverage vector} $\xx_i \in \calC_{k_i}$ (where $\calC_k= \{ x\in [0,1]^m \vert \sum {x_t} \le k \}$).  Here, $x_{i,t}$ is the probability that target $t$ is protected by one of defender $i$'s security resources.\footnote{Provably, any such coverage vector can be implemented by a distribution over deterministic allocation strategies, each employing at most $k$ resources.} We denote by $\mathbf{X}=(\xx_i)_{i\in\calN}$. 



If the defenders are uncoordinated, the probability that target $t\in \calT$ is protected 
is 
\[
\textstyle c_t = \cov_t(X) = 1-\prod_{i\in \calN} (1-x_{i,t}).
\]
The \emph{overall coverage vector} $\cc = (c_t)_{t\in \calT}$ is assumed to be known to the attacker, which chooses its action after the defenders have committed to their distribution. 
Thus, 
the attacker and defenders' utilities upon an attack at $t$ are: 
\begin{gather}
\label{eq:u-a-t}
U^a(\cc,t)=U^a(c_t,t)=(1-c_t)\cdot{r^a(t)}+c_t\cdot{p^a(t)}\\
\label{eq:u-d-t}
U_i^d(\cc,t)=U_i^d(c_t,t)=c_t\cdot{r_i^d(t)}+(1-c_t)\cdot{p_i^d(t)}
\end{gather}
In the classic (non-robust) model, all players are assumed to be rational. As such, the attacker's best response is to attack a target in the set
$\textstyle \BR(\cc) := \arg\max_{t\in \calT} U^a(\cc, t)$, 
which maximizes its expected utility. However, as discussed in the introduction, $\textstyle \BR(\cc)$ typically consists of multiple targets, which brings about the issues of tie-breaking and discontinuity. 

\paragraph{The Robust Model.}
The non-robustness of the standard MSSG model arises from the assumption of exact deterministic behavior of the attacker, whereby it always plays the \emph{exact} optimal play, even if the difference between the optimal play and the next in line is minuscule. Therefore, small changes in the attacker's utility function may lead to abrupt changes in the attacker's strategy, in turn causing abrupt changes in the defender's utilities. Accordingly, to obtain a robust model, we model the attacker's behavior as being \emph{probabilistic}, with a continuous distribution concentrated around the deterministic behavior dictated by the presumed (possibly inaccurate) utility. Importantly, we seek a general model which can capture the multitude of possible reasons for the mentioned perturbations. The formal details follow.

Given a coverage vector $\cc$, and the resultant attacker's utility vector $\uu=\uu(\cc)=(U^a(\cc,t))_{t\in\calT}$ (over the different targets), the attacker's actual behavior is assumed to be determined by a probability distribution $\omega(\uu)\in [0,1]^m$, specifying the probability that the attacker will actually attack each target. The function $\omega$ is termed the \emph{Attacker's Behavior Function}, and is assumed to have the following properties:
\begin{Definition}[\bf Attacker's Behaviour Function (ABF)]
\label{def:ABF}
A continuously differentiable function $\omega: \mathbb{R}_+^m \rightarrow \mathbb{R}_+^m$ is an \emph{attacker's behaviour function} if the following axioms hold:
\begin{enumerate}
    \item \label{axiom:abf-prob-dist}
    $\omega(\uu)$ is a probability distribution.
    \item \label{axiom:abf-monotonicity}
    $\omega$ is monotone increasing at each coordinate.
    \item \label{axiom:abf-independence}
    For each $ \emptyset \neq S \subseteq\calT$ and $t\in S: \frac{\omega_t(\uu)}{\sum_{t'\in S}{\omega_{t'}(\uu)}}$ is independent of any $u_{t'}$, $t'\not \in S$.
\end{enumerate}
\end{Definition}
Axiom~\ref{axiom:abf-independence} states that, given that the attack is within the set $S$, the conditional probability of an attack on any specific target $t\in S$ is only determined by the inter-relationships between the utilities of targets within $S$.

Thus, given a coverage vector $\cc$, inducing attacker utilities $\uu(\cc)$, and ABF $\omega$, defender $i$'s expected utility is given by:
\begin{equation}
\label{eq:u-d}
U_i^d(\cc) = \sum_{t\in \calT}  U_i^d(\cc,t)\omega_t(\uu(\cc))
\end{equation}
The following definition provides that the attacker's behavior is centered around the optimal deterministic one. 
\begin{Definition}[\bf $(\mathbold{\delta},\mathbold{\epsilon})$-ABF]
\label{def:ed-ABF}
Let $G$ be an (M)SSG, and $\delta,\epsilon>0$. An ABF $\omega$ is a $(\delta,\epsilon)$-ABF if:
\begin{equation}
\label{axiom:abf-sensitivity}
u_{t'}<u_t-\delta \Rightarrow \omega_{t'}(\uu)<\epsilon \\
\end{equation}
\end{Definition}
Thus, with a $(\delta,\epsilon)$-ABF, any target offering a utility $\delta$ less than the optimal one will be attacked with probability $<\epsilon$.
In Appendix~\ref{sec:ABFexample}, we show that one example of an ABF is the softmax function $\sigma(x)$, and that for any $\delta,\epsilon$, there exists a constant $d$, such that $\sigma(d\cdot)$ is a $(\delta,\epsilon)$-ABF. 

We believe this $(\delta,\epsilon)$-ABF formulation captures, under one unifying definition, the many possible sources for noise and uncertainty in the attacker's utility, as discussed in the introduction.
With these definitions, a \emph{robust MSSG} is a pair $\GG=\langle G, \omega \rangle$, where $G$ is a MSSG and $\omega$ is a $(\delta,\epsilon)$-ABF.\footnote{We note that inevitably the proper values for $\delta,\epsilon$ are dependent on $G$.  The reason is that $\delta$ must be small compared to the attacker's utilities, thus must be scaled in accordance, and $\epsilon$ must depend on the number of targets as it has to be small relative to $1/m$.}





\section{Solution Concepts}
For completeness, we review the MSSG solution concepts, as defined in previous works. Throughout, we consider the approximate versions.


\paragraph{$\mathbold{\zeta}$-Nash Equilibrium.}
Given the players' utility functions, the definitions of approximate NE is standard: 
\begin{Definition}[\bf $\mathbold{\zeta}$-Nash Equilibrium]
A strategy profile $\mathbf{X}=(\xx_i)_{i\in \calN}$ is an $\zeta$-NE if for any defender $i \in \calN$ and any strategy $\xx_i'$ of $i$:
\begin{equation}
\label{eq:NE-SSG}
U_i^d(\cov(X)) \ge U_i^d(\cov(\langle \xx_i', X_{-i} \rangle))-\zeta.
\end{equation}
\noindent
\end{Definition}


\paragraph{Coalitions.}
When the defenders are uncoordinated, independent choices may result in inefficient resource use. Therefore, \cite{mutzari2021coalition} consider a model for coalition formation in MSSG, which we adopt. Any subset $P\subseteq \calN$ of defenders may form a \emph{coalition}, in which case they act as a single defender with $k_P = \sum_{i \in P} k_i$ resources.
The coalition consisting of all of the defenders is called the {\em grand coalition}. 

Coalitions partition the set of defenders:
$\calN = \{P_1, \dots, P_\ell\}$. 
Each coalition $P_i$ chooses a coverage vector  $\xx_j \in \calC_{k_{P_j}}\}$.
Denoting $\mathbf{X} = (\xx_1, \dots, \xx_\ell)$, the probability that target $t$ is protected is:
\begin{equation}
\label{eq:coverage-product}
\textstyle c_t=\cov_t(\mathbf{X}) := 1-\prod_{i= 1}^{\ell}\left( 1- x_{i,t} \right).
\end{equation}
Given a strategy profile $\mathbf{X}$ and a target $t$, the defenders' utilities are defined in the same way as in \eqref{eq:u-d}. A {\em coalition structure} is any such pair $\CS \langle \calP, \mathbf{X} \rangle$.


Since $\CS$ includes a coverage vector $\mathbf{X}$ that (by \eqref{eq:coverage-product}) induces the overall coverage vector $\cov(X)$, for notational simplicity, the players' utilities can be viewed as a function of the coalition structure, which we simply denote $U^d_i(\CS)$. 

\paragraph{Deviations.}
Given a coalition structure $\CS$, a subset of defenders may choose to deviate and form a new coalition.  
In such a case, one must define the strategy played by all defenders. 
Following \cite{MOULIN1982115,chalkiadakis2011computational,shapley1973game}, the assumption is that in order to protect the status quo, the remaining defenders take revenge against the deviators, employing a strategy that makes at least one not gain by the deviation. It is assumed that this revenge strategy is chosen \emph{after} the deviators choose their strategy. The deviation is \emph{successful} if no such revenge is possible. 

\begin{Definition}[\bf $\mathbold{\zeta}$-successful deviation]
A deviation $(D,\xx_D)$ from a coalition structure $\CS$ is \emph{$\zeta$-successful} if for any deviator $i \in D$ and for any revenge strategy $\xx_R$, $U_i^d(\cov(\langle \xx_D, \xx_R \rangle)) > U_i^d(\cov(\CS)) + \zeta $.
\end{Definition}

\begin{Definition}[\bf $\mathbold{\zeta}$-approximate core]
A coalition structure $\CS$ is in the \emph{$\zeta$-approximate $\alpha$-core} if it admits no $\zeta$-approximate successful deviation.
\end{Definition}
So, in the $\zeta$-approximate core, a deviation cannot add more than $\zeta$ to the utility of at least one of the deviation's members.



\section{The Robustness Theorem}
We now provide the central robustness theorem, which essentially states that in the robust MSSG model, any approximate-NE or approximate-core {automatically} translates to their respective \emph{robust} counterparts.  We note that, technically, the theorem essentially follows directly from the continuity of $\omega$, but the result is exactly what is necessary. 

First, recall that the definition of a robust solution (Definition \ref{def:robust-solution}) requires a distance function between games. In MSSG, for $G=(\calN,\calT,\mathcal{K}, \mathcal{R},\calP), \hat{G}=(\calN,\calT,\mathcal{K}, \hat{\mathcal{R}},\hat{\calP})$, and $\GG=(G,\omega),\hat{\GG}=(\hat{G},\hat{\omega})$, we define:
\begin{align*}
d(\GG,&\hat{\GG})=
&\max \{ \| \mathcal{R}-\hat{\mathcal{R}} \|_{\infty},\| \calP-\hat{\calP} \|_{\infty},
\| \omega - \hat{\omega} \|_{\infty}
\}
\end{align*}
The following lemma bounds the change in utility arising from small changes in the game parameters. 
\begin{Lemma}
\label{lemma:robustness}
Let $\GG$ be a robust MSSG and let $\eta>0$. Then there exists a constant $b$ such that, for any coverage vector $\cc$ and any game $\hat{\GG}$ $\eta$-near to $\GG$, the following holds:
\begin{align*}
|\hat{U}_i^d(\cc) - U_i^d(\cc)| \le \frac{b}{2}\eta,
\end{align*}

where $\hat{U}_i^d(\cc)$, $U_i^d(\cc)$ are the utilities of defender $i$ in games $\hat{\GG}$, $\GG$ respectively.
\end{Lemma}
The proof, which essentially follows from the continuity of the ABF is provided in the Appendix~\ref{sec:proof-lemma-robustness}.

\begin{Theorem}
\label{theorem:delta-robustness}
Let $\GG$ be a MSSG. With $b$ of Lemma \ref{lemma:robustness}, for any $\eta$:
\begin{enumerate}
    \item any $\zeta$-NE of $\GG$ is also an $\eta$-robust $(\zeta+b\eta)$-NE of $\GG$.
    \item any $\zeta$ core of $\GG$ is also an $\eta$-robust $(\zeta+b\eta)$-core of $\GG$ (for all three variants of the core $\alpha,\beta,\gamma$).
\end{enumerate}  
\end{Theorem}
\begin{proof} We prove (1), and the proof of (2) is analogous. Let $\mathbf{X}$ be a $\zeta$-NE of $\GG$, and $\hat{\GG}$ $\eta$-near to $\GG$. Then, for any possible $\xx'_i$ of player $i$
\begin{align}
\hat{U}_i^d(\cov(\langle \xx'_i,\mathbf{X}_{-i}\rangle))& \ge {U}_i^d(\cov(\langle \xx'_i,\mathbf{X}_{-i} \rangle))-  b\eta/2 \label{eq:first}\\
&\geq  U_i^d(\cov(\langle \xx_i, \mathbf{X}_{-i} \rangle)) - \zeta - b\eta/2 \label{eq:second}\\
&\geq  \hat{U}_i^d(\cov(\langle \xx_i, \mathbf{X}_{-i} \rangle))-\zeta - b\eta 
\label{eq:third}
\end{align}
where we use Lemma \ref{lemma:robustness} for \eqref{eq:first} and \eqref{eq:third}, and \eqref{eq:second} follows by the definition of $\zeta$-NE.
\end{proof}
Accordingly, in order to find robust solutions in our model it suffices to find regular solutions in the model, and robustness then follows automatically. The remainder of the paper is thus dedicated to finding such solutions.

\section{Nash Equilibrium in Robust MSSGs}
We now show how to compute an approximate NE in robust MSSGs. For simplicity, we restrict the discussion to \emph{non-saturated} games.  
A robust MSSG $\GG$ is \emph{$\alpha$-saturated} if there exists a coverage $\cc\in \calC_{k_{\calN}}$ and $t \in \calT$ s.t. $c_t \ge 1-\alpha$ and $\omega_t(\cc) \ge \epsilon$.  In a saturated game, there are sufficient resources to induce an attack that is almost surely caught.
The results can be extended to saturated games using a similar approach to the one used by~\cite{gan2018stackelberg}.
\begin{Theorem}
\label{theorem:eps-NE-construction}
There exists a polynomial algorithm, such that for any $G$ there exist $A,B,C,\epsilon_0,\delta_0$ s.t. for any $(\delta,\epsilon)$-ABF $\omega$ with $\epsilon < \epsilon_0, \delta<\delta_0$, on input $(G,\omega)$ the algorithm outputs a strategy profile $\mathbf{X}$ that is a $\zeta$-NE, for $\zeta=B\delta + C\epsilon$ (provided $(G,\omega)$ is not $A\delta$-saturated).
\end{Theorem}
The exact constants, together with a detailed proof, appear in the Appendix~\ref{sec:ne}. The proof builds upon the techniques of~\cite{gan2018stackelberg}, but requires significant adaptations for our setting. For simplicity, in the main body of the paper we consider the  case where all targets carry identical penalties and identical rewards for the attacker (but not the defenders).  This case allows us to explain the core elements of the construction and the proof, while avoiding the technicalities. Note that in the explanation we do not seek to obtain the best constants. Better bounds are offered in the complete proof. 

For the case we are considering, we can further normalize the attacker's utilities so that $p^a(t)=0$ and $r^a(t)=1$ for all $t$.  With this assumption, $U^a(c,t)=1-c_t$, for all $c,t$.   

The core procedure underlying the algorithm is \textsc{ALLOC} (Algorithm \ref{alg:sim-alloc}). 
\begin{algorithm}[tb]
\caption{ALLOC}
\label{alg:sim-alloc}
\textbf{Input}: value $\parmc,\tildec\in [0,1]$, $G=(\calN,\calT,\mathcal{K},\mathcal{P},\mathcal{R})$\\
\textbf{Output}: a strategy profile $\mathbf{X}=( x_{i,t})$ \\
\begin{algorithmic}[1]
\STATE $c_t \leftarrow 0$, for all $t$
\FOR{$i \in \calN$}
\FOR{$t \in \calT$ by reverse $\preceq_i^{\tildec}$ precedence order} 
\STATE $x_{i,t} \leftarrow \min\{1-\frac{1-\parmc}{1-c_t}, k_i\}$
\STATE $k_i\leftarrow k_i-x_{i,t}$ 
\STATE $c_t \leftarrow 1-(1-c_t)(1-x_{i,t})$;
\ENDFOR
\ENDFOR  
\end{algorithmic}
\end{algorithm}
\textsc{ALLOC} gets as input a parameter $\parmc$, and aims to have $c_t=\parmc$, for all targets.  To this end, the defenders --- one by one in order --- iterate through the targets, allocating resources until either (i) the $\parmc$ level is reached, or (ii) their resources are fully depleted. This allocation is performed in Line 4, where $x_{i,t}=1-\frac{1-\parmc}{1-c_t}$ brings the coverage to exactly $\parmc$, and $x_{i,t}=k_i$ depletes the defender's resources.  Importantly, each defender considers the targets in \emph{reverse preference order} of its expected utility from the target, assuming some identical coverage level $\tildec$ (with $\tildec$ possibly different from $\parmc$) . Specifically, for  targets $t,t'$, we denote $t\preceq^{\tildec}_i t'$ if $U^d_i(\tildec,t)\leq U^d_i(\tildec,t')$.  In \textsc{ALLOC}, defenders iterate through the targets from the least to the highest in the $\preceq^{\tildec}_i$ order.

Clearly, the actual coverage level obtained by \textsc{ALLOC} is determined by $\parmc$: if $\parmc$ is too small then \textsc{ALLOC} may complete without depleting the defenders' resources, and if $\parmc$ is too big then some targets my have $c_t<\parmc$. However, there exists a $\parmcopt$ for which the algorithm completes with $c_t=\parmcopt$ for all targets (assuming the game is non-saturated). This $\parmcopt$ can be approximated to within any level of accuracy by binary search, with multiple runs of \textsc{ALLOC}. 
Set $\beta=(1-\parmcopt)/2$.

Set $\barc=\parmcopt+m\delta/\beta$. For $\delta$ sufficiently small, $\barc<1-\beta$.  Consider the outcome of running \textsc{ALLOC} with $\parmc=\barc$ and $\tildec=\parmcopt$. Let $\hat{\mathbf{X}}$ be the resultant allocation, and for each $t$, let $\hat{c}_t$ be the resultant coverage of target $t$. Since $\barc > \parmcopt$, there will necessarily be targets $t$ for which $\hat{c}_t<\parmcopt$, but for $\delta$ sufficiently small there will be only one such target. Denote this target $t^*$. For all other $t$'s, $\hat{c}_t=\barc$.

We now argue that $\hat{\mathbf{X}}$ is a $\zeta$-NE for $\zeta=B\delta + C\epsilon$, for some constants $B,C$ dependant only on $G$. 

In the following, we focus on the attacker's behavior occurring with probability $\geq \epsilon$.  For ease of exposition, we say that an event is \emph{likely} if it happens with  probability $\geq \epsilon$. 

If all defenders play by $\hat{\mathbf{X}}$ then an attack is only likely at $t^*$. This is since $\hat{c}_t-\hat{c}_{t^*}\geq \delta$, and hence $\hat{u}_{t^*}-\hat{u}_{t}\geq \delta$, for all $t$, and $\omega$ is a $(\delta,\epsilon)$-ABF.  
Consider a deviation 
of defender $i$, and let $\cc'$ be the coverage vector resulted from this deviation. 
We now explain why $\mathbf{c}'$ cannot ``substantially'' increase the utility of $i$. Let: $L$ be the targets - aside from $t^*$ to which $i$ allocated resources, $L^+=L\cup \{ t^*\}$, and $H=\calT\setminus L^+$.  We consider what $\mathbf{c}'$ can do to attacks on $H, L$ and $t^*$. 

\paragraph{Attacks on $H$.} We argue that under $\cc'$ the attacker is unlikely to attack any target of $H$. The reason is that since $i$ did not allocate resources to $H$, the only way that it can induce an attack on $H$ is by increasing the coverage of all targets in $L^+$ to $\barc+\delta$. But, since $c_{t^*}<\parmcopt=\barc-m\delta/\beta$, bringing $t^*$'s coverage to $\barc+\delta$ can only be accomplished by taking coverage from the members of $L$. In doing so, at least one will result with coverage less than $\barc-\delta$ (the value of $\barc$ was so chosen).

\paragraph{Attacks on $L$.}
A deviation \emph{can} induce attacks on $L$. However, it cannot provide $i$ substantially more than it originally obtained, where \emph{substantially} means adding more than $\delta B$ utility. The reason is that since $t^*$ is not $\barc$ covered, it must be that $t^*$ was considered by $i$ after all elements of $L$ - or else $i$ would either bring $t^*$'s coverage to $\barc$ or fully deplete its resources. So, $t\preceq^{\parmcopt}_i t^*$, for all $t\in L$. This means that - at coverage level $\parmcopt$ - attacks on targets of $L$ offer $i$ no more utility than attacks on $t^*$. So, if $\parmcopt$ where the coverage, then inducing an attack on $L$ would offer no gain to $i$. In practice, the elements of $\mathbf{c}'$ and $\hat{\mathbf{c}}$ are not exactly $\parmcopt$, but they are $O(\delta)$ away.  So, since $U^d_i(c,t)$ is linear in $c$, the differences in utility can also be only $O(\delta)$. So attacks on $L$ cannot offer \emph{substantially} more utility than what $t^*$ initially offered. 

\paragraph{Attack on $t^*$.}
A deviation can also possibly increase the coverage of $t^*$, thus offering more utility if and when attacked. However, one can only add $O(\delta)$ coverage (while keeping an attack on $t^*$ likely), so that this addition cannot add more than $O(\delta)$ utility.

\paragraph{Putting It All Together.} 
We obtain that with probability $1-\epsilon$ the added utility due to the deviation is at most $\delta B$, for some constant $B$.  With probability $\epsilon$ the increase can be larger, but clearly bounded by $C=\max_{j,t}{r^d_j(t)}$.  So, the utility increase is bounded by $\zeta = \delta B+ \epsilon C$. This completes the construction of $\zeta$-NE.  By Theorem \ref{theorem:delta-robustness}, this strategy is also an $\eta$-robust $(\zeta+b\eta)$-NE, for any $\eta$. 

\section{The Robust Core}
In this section we explain the algorithm to construct a robust $\zeta$-approximate $\alpha$-core. Importantly, we only outline the construction and proof ideas. Full details appear in Appendix~\ref{sec:app-core}.



\paragraph{Resistance to Subset Deviations.}
With minor adaptations, the ALLOC procedure can be applied to the cooperative setting (wherein probabilities are additive rather than multiplicative). Also, the procedure can easily be configured to accept a target utility level $u$ as input (rather than target coverage $c$) - see Algorithm~\ref{alg:gc-alloc}.  As before, repeated calls to ALLOC allow us to find a $\uopt$ and strategy $\bXX$ such that all targets offer utility $\uopt$ to the attacker (except for those with $r^a(t)<\uopt$), using all resources. Next, we re-run ALLOC with $\highu=\uopt-mO(\delta)$. 
Let $\hXX$ be the resulted strategy profile, $\hcc=\cov(\hXX)$, and $t^*$ the target not covered to $\highu$. Then, $U^a(\hcc,t^*)>\uopt + \delta$ and $U^a(\hcc,t)=\highu$ for all other $t$.  So, the attacker is only likely to attack $t^*$.  

We argue that $\hat{\mathbf{X}}$ is resilient to any deviation $\xx_D$ of any proper subset $D\subset \calN$ of defenders. Set $\cc'=\cov(\hXX_{-D},\xx_D)$. Let $L$ be the targets, aside from $t^*$, to which $D$ allocated resources in $\hXX$, and $L^+=L\cup\{ t^*\}$. Then, analogously to the NE case, no deviation of $D$ can induce a likely attack outside $L\cup\{ t^*\}$. They may, however, be able to alter the attack probabilities within $L^+$. Let $\mathcal{A}$ be the targets that are likely 
to be attacked under $\xx_D$. So, $|U^a(\cc',t_i)-U^a(\cc',t_j)|<\delta$, for any $t_i,t_j\in \mathcal{A}$.  Now, recall that given $D$ and $\xx_D$, the remaining defenders - $\bar{D}$ - can change their strategy. So, using only $O(\delta)$ resources per target, $\bar{D}$ can raise the coverage of all but one target of $t_0\in \mathcal{A}$, so that $t_0$ offers at least $\delta$ more utility than \emph{all} other targets. Since $t_0\in L^+$, there exists at least one $i\in D$ for which $t_0\preceq_i^{\cov(\XX)}t^*$. So, for this $i$, the deviation does not offer any (substantial) gain.  

\paragraph{Resistance to Grand Coalition Deviations.}
We now show how transform the above strategy to one resistant to grand coalition deviations.
Let $\hu^d_i$ be the utility of defender $i$ under $\hXX$. Consider the following linear program (with variables $p_t$):
\begin{align*}
\mathbf{max} &\sum_{t\in \calT}\sum_{i\in \calN} p_t \cdot U^d_i(\hcc,t)&\\
 s.t. &\sum_{t\in \calT} p_t\cdot U^d_i(\hcc,t)\geq \hu^d_i,& i=1,\ldots,n\\
& \sum_{t\in \calT} p_t=1; p_t \geq 0, &t=1,\ldots, m
\end{align*}
Let $\spp=(\ssp_t)_{t=1}^m$ the solution provided by the programs (there necessarily exists a solution, since $\pp=\omega(\uu(\hcc))$ is a feasible one). 
The following proposition (proved in Appendix~\ref{sec:proof-abf-surjectivity}) provides that with small changes to $\mathbf{\uopt}=(\uopt,\ldots,\uopt)$ the distribution $\spp$ can (essentially) be induced.
\begin{Proposition}
\label{prop:abf-surjectivity}
For any probability distribution $\pp$ and any $u>0$, there exists a utility vector $\uu$ s.t. $\|\omega(\uu)-\pp\|_\infty<  m^2\epsilon$ , and $u \le u_t < u + 2m\delta$ for all $t\in\calT$.
\end{Proposition}
Let $\tsuu$ be the utility vector provided by invoking Proposition \ref{prop:abf-surjectivity} with $u=\uopt$ and $\pp=\spp$. Since $\mathbf{\uopt}$ is implementable (by $\bXX$) and $\uopt^*_t\geq \uopt$ for all $t$, then $\tsuu$ is also implementable, by some strategy $\tsXX$. We now argue that $\tsXX$ is an the $\zeta$-approximate $\alpha$-core, for $\zeta=A\epsilon+B\delta$ (for some constants $A,B$ fully determined in the full proof). In the following arguments, 
interpret all statements ``up to at most $\zeta$ gains''.

To see that $\tsXX$ resists grand coalition deviations, first note that we may assume that in any such deviation $\XX'$, all elements of $\uu'=\uu(\cov(\XX'))$ are close (within $O(\delta)$) to $\uopt$.  Otherwise, there must be targets that are more than $\delta$ apart in their utility, in which case it is possible to shift resources from the those with lesser $u'_t$ to the higher ones, without substantially changing the attack distribution (by Proposition~\ref{prop:abf-surjectivity}).
Now, any target where $u_t' > \uopt$ is not substantially better for any deviator, and any target where $u_t' < \uopt - \delta$ is not likely to be attacked. By construction, $\spp$ maximizes the sum of defender utilities, subject to each getting at least as in $\hXX$.   So, it is impossible that \emph{all} defenders simultaneously get even more.  

For subset deviations, consider $D\subset \calN$ and suppose they have a deviation $\xx_D$ that guarantees all members of $D$ more than in $\tsXX$ (regardless of how the others play). But $\tsXX$ offers essentially as much as the output of the linear program, which, by its constraints, offers each defender as much as $\hXX$. So, $\xx_D$ would also constitute a successful deviation from $\hXX$, which we proved cannot be.  We thus obtain:
\begin{Theorem}
\label{theorem:core-construction}
There exists a polynomial algorithm, such that for any $G$ there exist $A,B,C,\epsilon_0,\delta_0$ s.t. for any $(\delta,\epsilon)$-ABF $\omega$ with $\epsilon < \epsilon_0, \delta<\delta_0$, on input $(G,\omega)$ the algorithm outputs a strategy profile $\mathbf{X}$ that is a $\zeta$-approximate $\alpha$-core, for $\zeta=B\delta + C\epsilon$ (provided that $(G,\omega)$ is not $A\delta$-saturated). 
\end{Theorem}
By Theorem \ref{theorem:delta-robustness}, for any $\eta>0$, any such solution is also an $\eta$-robust $(\zeta+b\eta)$-approximate $\alpha$-core. 

\paragraph{$\mathbold{\gamma}$-Core.}
Unlike the $\alpha$-core, the approximate $\gamma$-core (see~\cite{chander2010cores}, Definition~\ref{def:gamma-core}) may be empty.

\begin{table}[ht!]
\centering
\begin{tabular}{ |p{1.9cm}|p{0.5cm}|p{0.5cm}|p{0.5cm}|p{0.5cm}|p{0.5cm}|p{0.5cm}||p{0.25cm}|}
\hline
Player   & $t_1$ & $t_2$ & $t_3$ & $t_4$ & $t_5$ & $t_6$ & k \\
\hline\hline
Defender 1,2   & $1,0$ & $1,0$ & $1,0$ & $6,5$ & $6,5$ & $6,5$ & 1 \\
Defender 3,4   & $6,5$ & $6,5$ & $6,5$ & $1,0$ & $1,0$ & $1,0$ & 1 \\
\hline
Attacker & $1,0$ & $1,0$ & $1,0$ & $1,0$ & $1,0$ & $1,0$ & -- \\
\hline
\end{tabular}
\caption{Example of a MSSG where the approximate $\gamma$-core is empty}
\label{table:empty-gamma-core}
\end{table}

Basically, defenders $1,2$ can always $\gamma$-deviate and get a utility $\ge 4$ each, as can defenders $3,4$, but no coalition structure can yield all defenders $\ge 4$ utility. See a full explanation in Appendix~\ref{example:empty-gamma-core}.

\bibliographystyle{named}

\newpage
\appendix
\noindent 
{\huge{\bf{Appendix}}}
\section{Proof of Lemma \ref{lemma:robustness}}
\label{sec:proof-lemma-robustness}
\begin{proof}
The change of utility can be bounded, using the triangular inequality, by three additive factors in the following manner:


\begin{align}
\abs{\hat{U}_i^d(\cc)-&\hat{U}_i^d(\cc)} = \nonumber \\
\abs{\sum_{t\in \calT} \hat{U}_i^d(c_t,t)&\cdot \hat{\omega}_t(\hat{\uu}(\cc)) -\sum_{t\in \calT} {U}_i^d(c_t,t)\cdot \omega_t({\uu}(\cc))}  \leq \nonumber\\
\abs{\sum_{t\in \calT} \hat{U}_i^d(c_t,t)& \hat{\omega}_t(\hat{\uu}(\cc)) -\sum_{t\in \calT} U_i^d(c_t,t) \hat{\omega}_t(\hat{\uu}(\cc))} \label{eq:change-U}  \\
+ \abs{\sum_{t\in \calT} U_i^d(c_t,t)& \hat{\omega}_t(\hat{\uu}(\cc)) - \sum_{t\in \calT} U_i^d(c_t,t) \omega_t(\hat{\uu}(\cc))}\label{eq:change-w}  \\
+ \abs{\sum_{t\in \calT} U_i^d(c_t,t)& \omega_t({\hat{\uu}}(\cc)) - \sum_{t\in \calT} {U}_i^d(c_t,t) \omega_t({\uu}(\cc))} \label{eq:change-u}  
\end{align}

Intuitively, each term corresponds to the contribution of the change defender $i$'s parameters, the attacker's behaviour function, and the attacker's reward and penalty parameters, to the overall change of utility.

We now bound each of \eqref{eq:change-U}-\eqref{eq:change-u}. For \eqref{eq:change-U}:
\begin{equation*}
\begin{aligned}
\sum_{t\in \calT} \abs{\hat{U}_i^d(c_t,t)-{U}_i^d(c_t,t))}\cdot \hat{\omega}_t(\hat{\uu}(\cc)) \le \sum_{t\in \calT} \eta \, \hat{\omega}_t(\hat{\uu}(\cc)) = \eta
\end{aligned}
\end{equation*}

For \eqref{eq:change-w}:

\begin{equation*}
\begin{aligned}
\sum&_{t\in \calT} U_i^d(c_t,t) \abs{\hat{\omega}_t(\hat{\uu}(\cc)) - \omega_t(\hat{\uu}(\cc))} \\
&\le \eta \sum_{t\in \calT}{U_i^d(c_t,t)}
\le \eta \sum_{t\in \calT} r_i^d(t)
\end{aligned}
\end{equation*}

For \eqref{eq:change-u} we must first understand how a change in the attacker's utility vector alters the attack distribution of the targets. Consider $\omega$ restricted to the domain is $\prod_{t\in\calT} [\max(0,p^a(t)-\eta),r^a(t)+\eta]$ which is closed and bounded and therefore compact. Therefore, since $\omega$ is continuously differentiable, $| \nabla \omega |$ is continuous and therefore bounded, say by some constant $K$ (in other words, $\omega$ is $K$-Lipschitz continuous).\footnote{Note that $K$ can depend on $\eta$, but we are only interested in $\eta$, which is small relative to the game parameters, so this is negligible.}

Thus, by the mean value theorem,
\begin{equation*}
\begin{aligned}
\|\omega(\hat{\uu}(\cc))-\omega(\uu(\cc))\|_\infty &\le \|\omega(\hat{\uu}(\cc))-\omega(\uu(\cc))\|_2 \\
&\le K \| \hat{\uu}(\cc) - \uu(\cc) \|_2 \\
&\le K\sqrt{m} \| \hat{\uu}(\cc) - \uu(\cc) \|_\infty 
\end{aligned}
\end{equation*}

Therefore, for \eqref{eq:change-u}:
\begin{equation*}
\begin{aligned}
\sum_{t\in \calT} U_i^d(c_t,t)\abs{\omega_t(\hat{\uu}(\cc)) - \omega_t({\uu}(\cc))}\\ 
\le \sum_{t\in \calT} U_i^d(c_t,t) K\sqrt{m} \| \hat{\uu}(\cc)) - {\uu}(\cc))\|_{\infty} \\ 
\le \sum_{t\in \calT}  U_i^d(c_t,t)K\sqrt{m} \eta \le K\sqrt{m}\sum_{t\in \calT} r_i^d(t) \eta
\end{aligned}
\end{equation*}

Putting it all together, we get:
\begin{align*}
    \abs{\hat{U}_i^d(\cc)-\hat{U}_i^d(\cc)} \leq \eta(1+(1+K\sqrt{m})\sum_{t\in\calT}{r_i^d(t)}))
\end{align*}
\end{proof}

\section{Approximate Nash Equilibrium}
\label{sec:ne}




In order to describe our construction in detail, we will first introduce some definitions and notations. First, since all defenders prefer targets to be protected, we expect that in NE the defenders would utilize all of their resources. We call such a strategy profile an \emph{efficient} strategy profile.

\begin{Definition}[\bf Efficient Strategy Profile]
A strategy profile $X$ is efficient if for each defender $i\in\calN$, $\sum_{t\in\calT}{x_{i,t}} = k_i$.
\end{Definition}

Indeed, if a strategy profile $X$ is not efficient, then a defender can increase the coverage of all targets and gain more utility. However, if a target is already fully covered, this may not be possible. Therefore, for simplicity, in the the paper we assume that the game is not $\delta/b$-saturated, for $b=\min_{t\in\calT}{r^a(t)-p^a(t)}$. This follows the ideas from \cite{mutzari2021coalition} (they assume the game is canonical, which is equivalent to assuming that the game is not (0,0) saturated).

\begin{Definition}[\bf Height of a Coverage]
\label{def:height}
The height of a coverage vector $\cc \in [0,1]^m$, denoted $\hgt(\cc)$, is the optimal attacker utility it yields, $\hgt(\cc):=\max_{t\in \calT} U^a(\cc,t)$.
\end{Definition}


\begin{Definition}[\bf Mini-Max Height]
\label{def:min-max-height}
We denote the min-max attacker height by $$\overline{u}=\min_{\cc \in \calC_{k_\calN}}{\hgt(\cc)}=\min_{\cc \in \calC_{k_\calN}}{\max_{t\in \calT}{U^a(\cc,t)}}$$
\end{Definition}

Note that in the uncoordinated case scenario, there might not exist any strategy profile $X$ where $\hgt(\cov(X))=\overline{u}$, because the lack of cooperation introduces inefficiencies.



Lastly, we formally define the \emph{preference profile} induced by a coverage vector, which captures the heterogeneous preferences of the defenders.

\begin{Definition}[\bf Induced Preference Profile]
\label{def:induced-preference-profile}
A preference profile $J$ induced by a coverage vector $\cc$ is a list of binary relations $\langle \preceq_i^J \rangle_{i \in \calN}$ such that $t_1 \preceq_i^J t_2$ iff $U_i^d(\cc,t_1) \le U_i^d(\cc,t_2)$. For any $\overline{p}<u<\overline{r}:=\max_{t\in\calT}{r^a(t)}$, we define $\overline{\cov}(u)$ to be the coverage vector such that on every target $t\in\calT$, $U^a(\overline{\cov}(u),t)=\min(u,p^a(t))$. Overloading notations, the preference profile induced at $u$ is $J(u):=J(\overline{\cov}(u))$.
\end{Definition}

We are now ready to describe ~\cite{gan2018stackelberg}'s procedure, $\textsc{ALLOC}$, for canonical games, depicted in Algorithm~\ref{alg:alloc}. The algorithm gets as an input an attacker utility value $u$. In a greedy fashion, each defender $i$ in his turn allocates resources to targets according to his induced preference profile $J(u)$, $t_{i1} \preceq_i t_{i2} \preceq_i \dots \preceq_i t_{im}$, covering each target so that the attacker's utility on it gets down to $u$, or until he runs out of resources.

The main observation of~\cite{gan2018stackelberg} for canonical games can be phrased as follows: there exists an input $u^*$ such that \textsc{ALLOC}$(u^*)$ outputs an efficient strategy profile $X$ and $\cov(X)=\overline{\cov}(u^*)$. They also describe how to find $u^*$ efficiently. We denote by $t^*$ the last target that is covered in \textsc{ALLOC}$(u^*)$.
Next, if the last target is covered up to height $u^*-A\zeta$ for $A$ being dependent regarding the game parameters only, then the resulting strategy profile $X$ is an $\zeta$-NSE, which means it's an $\zeta$-NE assuming pessimistic defender beliefs. Provably, this works because any of the following deviation attempts fail:
\begin{enumerate}
\item \label{dev:induce-uncovered-targets}
A defender cannot induce the attacker to attack targets to which he didn't allocate resources. This is because it would require him to overly cover all of the other targets, which would require too many resources for the efficiency of $X$. 
\item \label{dev:induce-covered-targets}
A defender can induce the attacker to attack a target he covered by reducing its coverage, but by construction such targets give him less utility than $t^*$.
\item \label{dev:increase-coverage}
A defender can increase the coverage of the target $t^*$, but it is possible to choose an $A$ such that this will not give him $>\zeta$ utility gain.

\end{enumerate}

\begin{algorithm}[tb]
\caption{ALLOC}
\label{alg:alloc}
\textbf{Input}: $u,\tilde{u} \in (\overline{p},\overline{r})$; \\
\textbf{Parameter}: a canonical SSG G; \\
\textbf{Output}: a strategy profile $X=\langle x_{i,t} \rangle$; \\
\begin{algorithmic}[1]
\STATE Set $x_{i,t} \leftarrow 0$, $\Delta_t \leftarrow \overline{cov}_t(u)$ and $c_t \leftarrow 0$ for all $i,t$;
\FOR{$i \in \text{G.}\calN$}
\FOR{$t \text{ in G.}J_i(\tilde{u}) = t_{i1},\dots,t_{im}$}
\STATE $x_{i,t} \leftarrow \min{\{\Delta_t, \text{G.}k_i-\sum_{t'\in\calT}{x_{i,t'}}\}}$;
\STATE $c_t \leftarrow 1-(1-c_t)(1-x_{i,t})$;
\STATE \algorithmicif\ {$c_t = 1$} \algorithmicthen\ {$\Delta_t \leftarrow 0$} \algorithmicelse\ {$\Delta_t \leftarrow \frac{\overline{cov}_t(u)-c_t}{1-c_t}$};
\STATE \algorithmicif\ {$x_{i,t} > 0$} \algorithmicthen\ {$t^* \leftarrow t$};
\ENDFOR
\ENDFOR
\STATE \textbf{return} $X, t^*$;
\end{algorithmic}
\end{algorithm}

Although using the same construction ensures that deviation attempts of type \ref{dev:induce-covered-targets} and \ref{dev:increase-coverage} still fail in our model, deviation type \ref{dev:induce-uncovered-targets} becomes possible, since equally covered targets can both be attacked with significant probability. The lack of pessimistic belief forces us to make some adaptations to~\cite{gan2018stackelberg} in order to make it work (see Algorithm~\ref{alg:calc-ne}).




\begin{algorithm}[tb]
\caption{$\text{CALC}_\zeta$-NE}
\label{alg:calc-ne}
\textbf{Parameter}: (1) $\delta,\epsilon > 0$;\\
\hspace*{17mm} (2) a $\delta$-canonical game $\GG=(G,\omega)$;\\
\textbf{Output}: a $\zeta$-NE $X=\langle x_{i,t} \rangle$, $\zeta$;\\
\begin{algorithmic}[1]
\STATE right $\leftarrow$ G.$\overline{r}$, left $\leftarrow$ G.$\overline{p}$;
\WHILE{right - left $>$ $\delta^2$}
    \STATE u $\leftarrow$ (left + right) / 2;
    \STATE X, t = ALLOC(u,u);
    \STATE \algorithmicif\ {$U_i^d$(X, t) $>$ u} \algorithmicthen\ {left $\leftarrow$ u} \algorithmicelse\ {right $\leftarrow$ u};
\ENDWHILE

\STATE $\zeta \leftarrow$ G.B * $\epsilon$ + G.C * $\delta$;
\STATE \textbf{return} ALLOC(u - G.A * $\delta$,u), $\zeta$;

\end{algorithmic}
\end{algorithm}

\subsection{Proof of Theorem~\ref{theorem:eps-NE-construction}}

In this subsection we formally proof Theorem~\ref{theorem:eps-NE-construction}.

\begin{proof}

We divide into cases. Let $\bar{p} := \max_{t\in \calT}{p^a(t)}$, and $\bar{t}$ will be some target with attacker penalty $\bar{p}$. In the first case, calling procedure ALLOC with input $(u<\bar{p}+\delta,\cdot)$ gives a level coverage. In this case, increasing the attacker utility on target $\bar{t}$ by $\delta$ will make any other target be attacked with probability $<\epsilon$, and hence $\bar{t}$ is attacked with probability $\ge 1-(m-1)\epsilon$. This is in contradiction to the assumption that the game is not $(\delta/b, \epsilon)$-saturated, for $b = \min_{t\in\calT} {(r^a(t)-p^a(t))}$.

Next, since $\cov(ALLOC(u,u))$ is continuous with $u$, there exists some height $u^* \ge \bar{p} + \delta$ for which $X^*=ALLOC(u^*,u^*)$ forms a level coverage $\cc^*=\cov(X^*)$ of height $u^*$. This means that for any target $t\in\calT$, if $u^* < p^a(t)$ then $c_t^*=0$, and otherwise $U^a(\cc^*, t)=u^*$. This $u^*$ can easily be found using binary search up to any fixed precision. 

For the second case, denote by $a = \max_{t\in\calT} {(r^a(t)-p^a(t))}$, and $g := \min_{t\in\calT} {(u^* - p^a(t))} \ge \delta > 0$. Note that $g \ge \delta$ only for this second case. Fix the following values for $A,B,C$:
\begin{itemize}
    \item $ A := \frac{4ma^2}{bg} $
    \item \begin{align}
        B:= &\max_{i\in\calN} {(r_i^d(t^*)-p_i^d(t^*))} \cdot \nonumber\\
    &(\frac{8ma^3 \sum_{t\in\calT} {(r^a(t)-p^a(t))}}{bg^2} + A) \nonumber
    \end{align} 
    \item $ C := \max_{i\in\calN, t\in\calT}{r_i^d(t^*)} + \max_{i\in\calN} { \sum_{t\in\calT} {r_i^d(t)}} $
\end{itemize}

We also fix $\epsilon_0 := 1/m$ and $\delta_0 := \frac{bg^2}{8ma^2}$ and assume that $\delta<\delta_0$ and $\epsilon < \epsilon_0$. Consider $X=ALLOC(u^* - A\delta,u^*)$ and denote by $\cc=\cov(X)$ and by $\uu=(U^a(\cc,t))_{t\in\calT}$. We divide further into two cases, where in the first case, $g > A\delta$. For $\delta$ sufficiently small, there is only a single target $t^*$ that yields an attacker utility greater than $u^*$, which is also the last target covered in ALLOC($u^*$,$u^*$). For any other target $t$, if $p^a(t) > u^*-A\delta$ then $c_t=0$, and otherwise $u^t=u^*-A\delta$. We argue that $X$ is a $\zeta$-NE for $\zeta=B\delta+C\epsilon$ for some constants $B,C$ depdant only on $G$.

Since $A>1$, this means that with probability $\ge 1-m\epsilon$, target $t^*$ is being attacked and each defender gets a utility of $U_i^d(c_{t^*}, t^*)$. Let $i \in \calN$ be some defender who deviates with some strategy $x_i'$. Let $X'= \langle X_{-i}, x_i' \rangle$ be the resulted strategy profile and $\cc', \uu'$ the resulted coverage vector and attacker utility vector respectively. Let $S$ be the subset of targets not covered by defender $i$ in $X$, that is, $t\in S \iff x_{i,t}=0$.  Note that $\BR(X)=\{t^*\}$ by construction. Also note that, the targets covered by defender $i$ are less favorable for him by construction at height $u^*$, than target $t^*$. That is $\forall t\not\in S: t \preceq_i t^*$, meaning $\forall t\not\in S: U_i^d(u^*,t) \le U_i^d(u^*,t^*) $.

If the target $t^*$ is the most favorable target for defender $i$ at height $u^*$, then the maximum defender $i$ can do is increase the coverage of target $t^*$ while keeping the probability the attacker will attack target $t^*$ large. Since $X$ is efficient, meaning utilizes all the resources of all defenders, we know that the resulted height of $X$ is at most $\hgt(X) \ge u^* -A\delta$, therefore we can bound the utility after such a deviation by $U_i^d(\cov(u^*-A\delta),t^*) + \calO(\epsilon)$. Formally:

\begin{equation}
\label{eq: NLE-advantage-bound}
\begin{aligned}
U_i^d(X') &\le U_i^d(\cov(u^*-A\delta),t^*) + \epsilon \sum_{t \in \calT}{r_i^d(t)} \\
& = U_i^d(X,t^*) + \frac{r_i^d(t^*)-p_i^d(t^*)}{r^a(t^*)-p^a(t^*)}(\delta'+A\delta)
+ \epsilon \sum_{t \in \calT}{r_i^d(t)} \\
& \le U_i^d(X) + \max_{t\in\calT} r_i^d(t) \epsilon + \frac{r_i^d(t^*)-p_i^d(t^*)}{r^a(t^*)-p^a(t^*)}(\delta' + A\delta) \\
&+ \epsilon \sum_{t \in \calT}{r_i^d(t)}
\end{aligned}
\end{equation}

Where $U^a(X,t^*) - u^* = \delta'$ and the second inequality follows directly from Axiom~\ref{axiom:abf-sensitivity} because $A\delta+\delta' > \delta$, which is indeed the case since we take $A>1$, since $a \ge b$ and $a \ge g$.

Next, assume that there are targets in $S \setminus \{t^*\}$ that are more favorable for defender $i$. We will show that for any strategy he takes, in the resulting coverage vector $\cc'$, there will be at least one target $t' \not\in S$ such that for every target $s\in S$, $U^a(\cc',t') > U^a(\cc',s) + \delta$, and hence all targets in $t'$ will be attacked with probability at most $\epsilon$.

Indeed, assume the contrary. If defender $i$ tries to minimize
$$\Delta(\cc') := \min_{t\not\in S,s\in S} {U^a(\cc',t) - U^a(\cc',s)},$$
\noindent
he will allocate all of his resources to $\calT \setminus S$, resulting with equal attacker utility on all of those targets, utilizing all of his resources. We will denote this utility by $u_=$. Indeed, otherwise, he could simply shift resources from the more protected targets to the less ones in $\calT\setminus S$, and reduce $\Delta$. Therefore, we know that:

\begin{enumerate}
    \item $ \sum_{j\not\in S} x_{ij} = \sum_{j \not\in S} x_{ij}' = k_i $
    \item $ \forall j \in \calT \setminus \{t^*\}: 1 - (1 - x_{ij})(1 - c_{-j}) = \frac{r^a(j) - (u^*-A\delta)}{r^a(j) - p^a(j)} $
    \item $ 1 - (1 - x_{i,t^*})(1 - c_{-t^*}) = \frac{r^a(t^*) - (u^*+\delta')}{r^a(t^*) - p^a(t^*)} $
    \item $ \forall j \not\in S: 1 - (1 - x_{ij}')(1 - c_{-j}) = \frac{r^a(j) - u_=}{r^a(j) - p^a(j)} $
\end{enumerate}

We want to claim that $u^*+A\delta - u_= > \delta$. Let's simplify:

\begin{enumerate}
    \item $ \sum_{j\not\in S} x_{ij} = \sum_{j \not\in S} x_{ij}' = k_i $
    \item $ \forall j \in \calT \setminus \{t^*\}: 1 - x_{ij} = \frac{1}{1 - c_{-j}} \cdot \frac{u^*-A\delta - p^a(j)}{r^a(j) - p^a(j)} $
    \item $ 1 - x_{i,t^*} = \frac{1}{1 - c_{-t^*}} \cdot \frac{u^*+\delta' - p^a(t^*)}{r^a(t^*) - p^a(t^*)} $
    \item $ \forall j \not\in S: 1 - x_{ij}' = \frac{1}{1 - c_{-j}} \cdot \frac{u_= - p^a(j)}{r^a(j) - p^a(j)} $
\end{enumerate}

Therefore, subtracting $\cc-\cc'$, we get that:

\begin{enumerate}
    \item $ \sum_{j\not\in S} x_{ij} = \sum_{j \not\in S} x_{ij}' = k_i $
    \item $ \forall j \not\in S \setminus \{t^*\}: x_{ij}-x_{ij}' = \frac{1}{1 - c_{-j}} \cdot \frac{u_= - u^* + A\delta}{r^a(j) - p^a(j)} $
    \item $ x_{i,t^*} - x_{i,t^*}' = -\frac{1}{1 - c_{-t^*}} \cdot \frac{u^*+\delta' - u_=}{r^a(t^*) - p^a(t^*)} $
\end{enumerate}

Next, summing up all the equations in $2$ and equation $3$, and substituting equation $1$, we get:

\begin{equation}
\begin{aligned}
\sum_{j \in \overline{S} \setminus \{t^*\}}{ \frac{1}{1-c_{-j}}\cdot \frac{u_= - u^* +A\delta}{r^a(j)-p^a(j)}} \\
- \frac{1}{1-c_{-t^*}}\cdot \frac{u^*+\delta' - u_=}{r^a(t^*)-p^a(t^*)} = k_i-k_i = 0
\end{aligned}
\end{equation}

Next, note that $\Delta = u_= - (u^*-A\delta)$. Therefore:

$$ \Delta \cdot \sum_{j \not\in S}{ \frac{1}{1-c_{-j}}\cdot \frac{1}{r^a(j)-p^a(j)}} = \frac{1}{1-c_{-t^*}}\cdot\frac{\delta'+A\delta}{r^a(t^*)-p^a(t^*)} $$

Therefore, simplifying we get:

$$ \Delta \ge \frac{b(\delta'+A\delta)}{a(1-c_{-t^*}) \cdot \sum_{j \not\in S} \frac{1}{1-c_{-j}}} $$

Next, we remember that $0 \le c_{-j} \le \frac{r^a(j)-\overline{u}+A\delta}{r^a(j)-p^a(j)}$, since we covered up to height $u'=\overline{u}-A\delta$. Therefore, $1 \ge 1-c_{-j} \ge \frac{\overline{u}-A\delta-p^a(j)}{r^a(j)-p^a(j)}$, and we get $1 \le \frac{1}{1-c_{-j}} \le \frac{r^a(j)-p^a(j)}{\overline{u}-A\delta-p^a(j)} \le \frac{a}{g-A\delta} \le \frac{2a}{g}$ for $A\delta \le A\delta_0 = g/2$. Therefore, we get that:

$$ \Delta \ge \frac{bg(\delta'+A\delta)}{2a^2|T|} $$

Now we only need to verify $A\delta,\delta'$ are large enough so that $\frac{bg(\delta'+A\delta)}{2ma^2} > \delta $. We can do that by taking $A\delta = \frac{4ma^2}{bg} \delta \in \mathcal{O}(\delta)$. Now that $\Delta$ is at least $\delta$, targets outside of $S$ are attacked with probability at most $\epsilon$.

Lastly, for the case where $ g>A\delta $, we have to show that $\delta' \in \mathcal{O}(\delta)$ in order to bound the advantage of deviation. Intuitively, since the height $u^*+\delta'$ is a continuous function of the height of the rest of the targets, namely $u^*-A\delta$, if $A\delta$ is small enough then $\delta'$ is also small enough, therefore the utility loss from this NLE solution is $\mathcal{O}(\delta+\epsilon)$. We would want to give a concrete bound however, so let's look at the resource count.

In order to decrease an attacker utility by $A\delta$ one needs to increase the coverage of this target by $(r^a(t)-p^a(t))A\delta$. The cost of increasing the coverage by $(r^a(t)-p^a(t))A\delta$ from a value of $c_t$ by a single defender is $\frac{1}{1-c_t}(r^a(t)-p^a(t))A\delta \le \frac{2a(r^a(t)-p^a(t)}{g}A\delta$. However, a single defender may not have that amount of resource left by this stage. In worst case, we may assume that all of the defenders allocate this same amount of resource, which will always be enough to decrease the coverage by $A\delta$.

Therefore, the maximal resource loss we get from increasing the coverage to height $u^*-A\delta$ is $m\cdot \frac{2a\sum_t(r^a(t)-p^a(t))}{g}A\delta$, which will give an attacker utility decrease of $\delta' \le (r^a(t^*)-p^a(t^*))\frac{2am\sum_t(r^a(t)-p^a(t))}{g}A\delta$.

For the last part of the proof, we have that $\delta \le g < A\delta$. For this case, calling ALLOC with input $(u^*-A\delta, u^*)$, will lead to cases where some of the targets get fully covered and still the height $u^*-A\delta$ is not reached. Denote the set of those targets by $M$. Then by the assumption that the game is not saturated, by definition defender $i$ cannot induce an attack on targets in $S \cap M$, because they will stay fully covered. As for targets in $S \setminus M$, again defender $i$ will not have enough security resources to induce an attack on them. That is because now targets in $M \setminus S$ free him only less security resources to utilize.

\end{proof}

\section{Approximate \texorpdfstring{$\alpha$}{}-Core}
\label{sec:app-core}

\subsection{Proof of Resistance to Subset Deviations}
In this subsection we prove the first step of our construction: resistance to deviations of coalitions other than the grand coalition.
The adaption of ALLOC to the cooperative case is provided in Algorithm~\textsc{GC-ALLOC} is explicitly described in \ref{alg:gc-alloc}. 

\begin{algorithm}[tb]
\caption{GC-ALLOC}
\label{alg:gc-alloc}
\textbf{Input}: $u,\tilde{u} \in [\overline{p},\overline{r}]$; \\
\textbf{Parameter}: (1) a canonical game $\langle \rr^a,\pp^a,\rr^d,\pp^d,\kk \rangle$; \\
\hspace*{18mm}(2) $\delta,\epsilon > 0$ and a $(\delta,\epsilon)$-ABF $\omega$;\\
\textbf{Output}: a coalition structure $\CS=\langle \calN, X \rangle$ \\
\begin{algorithmic}[1] 
\STATE Set $x_{i,t} \leftarrow 0$, $\Delta_t \leftarrow \overline{cov}_t(u)$ and $c_t \leftarrow 0$, for all $i,t$;
\FOR{$i \in \calN$}
\FOR{$t=t_{i1},\dots,t_{im}$ in $J_i(\tilde{u})$}
\STATE $x_{i,t} \leftarrow \min{\{\Delta_t, k_i - \sum_{t'\in\calT}{x_{i,t'}}\}}$;
\STATE $c_t \leftarrow c_t+x_{i,t}$;
\STATE \algorithmicif\ {$c_t = 1$} \algorithmicthen\ {$\Delta_t \leftarrow 0$} \algorithmicelse\ {$\Delta_t \leftarrow \overline{cov}_t(u)-c_t$};
\ENDFOR
\ENDFOR
\STATE \textbf{return} $X$;
\end{algorithmic}
\end{algorithm}

\begin{Theorem}
\label{theorem:core-step1}
There exist $A,B,C,\delta_0,\epsilon_0>0$ s.t. for any $(\delta,\epsilon)$-ABF $\omega$ with $\epsilon<\epsilon_0$, $\delta<\delta_0$, the coalition structure $\CS$ generated by Algorithm \textsc{GC-ALLOC} with input $(u=\overline{u}-A\delta, \overline{u})$, is resistant to any deviation of a coalition $D \subset \calN$. That is, $D$ has no $\zeta=B\epsilon + C\delta$ successful deviation from $\CS$.
\end{Theorem}

\begin{proof}

We divide into cases. Let $\bar{p} := \max_{t\in \calT}{p^a(t)}$, and $\bar{t}$ will be some target with attacker penalty $\bar{p}$. In the first case, calling procedure ALLOC with input $(u<\bar{p}+\delta,\overline{u}$ gives a level coverage. In this case, increasing the attacker utility on target $\bar{t}$ by $\delta$ will make any other target be attacked with probability $<\epsilon$, and hence $\bar{t}$ is attacked with probability $\ge 1-(m-1)\epsilon$. This is in contradiction to the assumption that the game is not $(\delta/b, \epsilon)$-saturated, for $b = \min_{t\in\calT} {(r^a(t)-p^a(t))}$.

For the second case, denote by $a = \max_{t\in\calT} {(r^a(t)-p^a(t))}$, and $g := \min_{t\in\calT} {(u^* - p^a(t))} \ge \delta > 0$. Note that $g \ge \delta$ only for this second case. Fix the following values for $A,B,C,\delta_0,\epsilon_0$:
\begin{itemize}
    \item $A := 1 + \frac{ \sum_t (r^a(t) - p^a(t))^{-1}}{\min_t{(r^a(t) - p^a(t))^{-1}}}$
    \item $B := \max_{i\in\calN}{(r_i^d(t^*)-p_i^d(t^*))} \cdot [(r^a(t^*)-p^a(t^*))^{-1} + \sum_t{(r^a(t)-p^a(t))^{-1}} + \frac{(\sum_t{(r^a(t)-p^a(t))^{-1}})^2}{r^a(t^*)-p^a(t^*)}]$
    \item $C := \max_{i\in\calN} {\sum_{t\in\calT} {r_i^d(t)}}$
    \item $\delta_0 = (\sum_t {(r^a(t)-p^a(t))^{-1}})^{-1}$
    \item $\epsilon_0 = 1/m$
\end{itemize}

We divide further into two cases, where in the first case, $g > A\delta$. For $\delta$ sufficiently small, there is only a single target $t^*$ that yields an attacker utility greater than $\overline{u}$, which is also the last target covered in ALLOC($\overline{u}-A\delta$,$\overline{u}$). For any other target $t$, if $p^a(t) > \overline{u}-A\delta$ then $c_t=0$, and otherwise $u^t=\overline{u}-A\delta$. We argue that $X$ is a $\zeta$-NE for $\zeta=B\delta+C\epsilon$ for some constants $B,C$ dependant only on $G$.

In contradiction, let $(D,\cc_D')$ be an $\epsilon'$-successful deviation. Let $T_D$ be the set of targets such that $\cc_D(t)>0$, and $T_R=\overline{T_D}$. For each target $t \not\in T_D$, we fix the revenge takers coverage to the value that will give the attacker a utility of $\overline{u}-\delta$. As $\overline{u}$ is the mini-max height, this means that after the deviation targets on $\overline{T_D}$ will be attacked with probability at most $\epsilon$. Indeed, if the contrary is true, then the height of the coverage must be smaller than $\overline{u}$, the height of the mini-max coverage, a contradiction.

Next, observe the resulted coverage vector, and define $T_\epsilon$ to be the set of targets that have an attack probability of at least $\epsilon$. Note that $T_\epsilon \subseteq T_D$ as mentioned. Let $t_\BR \in \BR \subseteq T_\epsilon \subseteq T_D$ be some target.

Next, consider two cases. In the first, $|T_R| \ge 1$. In this case, the revengers will use the resources left to make any target in $T_D$ other than $t_\BR$ have an attacker utility smaller than $U^a(\cc_D',t_\BR)-\delta$. We will now verify that because of the choice of $A$, they have enough resources for the revengers to do that.

For each target $t\in T_R$, the coverage $c_R(t)$ required to get to an attacker utility $u$ is: $ c_R(t)=\frac{r^a(t)-u}{r^a(t)-p^a(t)} $.
Therefore, since $ (\overline{u}-\delta) - (\overline{u}-A\delta) = \frac{\sum_t (r^a(t) - p^a(t))^{-1}}{\min_t (r^a(t)-p^a(t))^{-1}}\delta$, the resources that the revengers save from decreasing the coverage of the targets in $T_R$ is:
$$ \Delta K_R^+ = \sum_{t' \in T_R} {(r^a(t')-p^a(t'))^{-1} \frac{\sum_t {(r^a(t) - p^a(t))^{-1}}}{\min_t {(r^a(t)-p^a(t))^{-1}}}\delta} $$
And since there is at least one target in $T_R$, it is lower bounded by:

\begin{equation}
\begin{aligned}
\Delta K_R^+ &\ge \min_{t'}{(r^a(t')-p^a(t'))^{-1} \cdot \frac{\sum_t (r^a(t) - p^a(t))^{-1}}{\min_t (r^a(t)-p^a(t))^{-1}}\delta} \\
&= \sum_t (r^a(t) - p^a(t))^{-1} \delta
\end{aligned}
\end{equation}

On the other hand, the amount of resource required to decrease all targets in $T_D \setminus \{t_\BR\}$ to have an attacker utility smaller than $U^a(\cc_D,t_\BR)-\delta$. However, this can be upper bounded by the cost of increasing all of the targets by $\delta$ attacker utility:
$$ \Delta K_R^- \le \sum_t (r^a(t)-p^a(t))^{-1}\delta $$
Therefore, $\Delta K_R^+ \ge \Delta K_R^-$, and there are enough security resources to do so.

In the second case, $T_R=\emptyset$, therefore $T_D=T$. This means that the deviators have allocated resources to every target. Then once again, if the revengers allocate resources to every target $t$ other than $t_\BR$ in order to make its attack probability less than $\epsilon$, this will require at most $\Delta K_R^- \le \sum_t {(r^a(t)-p^a(t))^{-1} \delta}$ resources, and therefore for $\delta \le \delta_0 = (\sum_t {(r^a(t)-p^a(t))^{-1}})^{-1}$, this requires $\Delta K_R^- \le 1$ resources, which we assume the revengers have.

In any case, we see that the revengers can induce the attacker to attack a single target with probability greater than $\epsilon$, and that this target is in $T_D$. However, consider a deviator $i\in D$ who allocated resources to target $t$.
Therefore, if we denote by $\CS'$ the coalition structure of the deviation and revenge coalitions described above, then the utility of defender $i$ can be bounded by:
$$ U_i^d(\CS') \le U_i^d(\CS',t_\BR) + \epsilon \sum_t r_i^d(t) $$
However, since $t_\BR \in \BR$ and $\overline{u}$ is the height of the level coverage, then target $t_\BR$ cannot be covered such that the attacker utility on it is lower than $\overline{u}$. Therefore, by construction and since defender $i$ covered target $t_\BR$ before target $t^*$, we have:
$$ U_i^d(\CS',t_\BR) \le U_i^d(\overline{\CS},t^*) $$
where $\overline{\CS}$ is a level coverage of height $\overline{u}-\delta$. Finally:
$$U_i^d(\overline{\CS}, t^*) = U_i^d(\CS, t^*) + \frac{r_i^d(t^*)-p_i^d(t^*)}{r^a(t^*)-p^a(t^*)}(\delta+\delta') $$

Where $\overline{u}+\delta'$ is the height of $\CS$, meaning the utility of the attacker on target $t^*$ in $\CS$. Putting this altogether, we get that:

$$ U_i^d(\CS')-U_i^d(\CS) \le \epsilon \sum_t{r_i^d(t)} + \frac{r_i^d(t^*)-p_i^d(t^*)}{r^a(t^*)-p^a(t^*)}(\delta+\delta') $$

All that is left is compute $\delta'$. We know that $\CS$ utilizes all $K$ resources, and so is a level coverage of height $\overline{u}$. Therefore:

\begin{equation}
\begin{aligned}
K &= \sum_t \frac{r^a(t)-\overline{u}}{r^a(t)-p^a(t)} \\
&= \sum_{t\neq t^*} \frac{r^a(t)-(\overline{u}-\delta')}{r^a(t)-p^a(t)} + \frac{r^a(t^*)-(\overline{u}+\delta')}{r^a(t^*)-p^a(t^*)}
\end{aligned}
\end{equation}

Simplifying this gives:

$$\delta' = \delta'(r^a(t^*)-p^a(t^*)) \sum_{t\neq t^*} (r^a(t)-p^a(t))^{-1} $$

Where $\delta' = \delta + \frac{\sum_t{(r^a(t)-p^a(t))^{-1}}}{\min_t{(r^a(t)-p^a(t))^{-1}}}\delta$

Therefore,

\begin{equation}
\begin{aligned}
\delta'/\delta &\le (r^a(t^*)-p^a(t^*)) \sum_t{(r^a(t)-p^a(t))^{-1}} \\
&+ (\sum_t{(r^a(t)-p^a(t))^{-1}})^2
\end{aligned}
\end{equation}

Thus, there are no $\zeta$ successful deviations where $K_D<D$, if

\begin{equation}
\begin{aligned}
\zeta & \ge \epsilon \sum_t{r_i^d(t)} & \\
& + \delta \max_i{(r_i^d(t^*)-p_i^d(t^*))} \cdot [ &  (r^a(t^*)-p^a(t^*))^{-1} + \\
& & \sum_t{(r^a(t)-p^a(t))^{-1}} + \\
& & \frac{(\sum_t{(r^a(t)-p^a(t))^{-1}})^2}{r^a(t^*)-p^a(t^*)}]
\end{aligned}
\end{equation}

This completes the proof for $g > A\delta$. Lastly, exactly by the same argument in the proof of Theorem~\ref{theorem:eps-NE-construction}, there are no $\zeta$ successful deviations in the case where $g \le A\delta$. 

\end{proof}

\subsection{Resistance to Grand Coalition Deviations}
In order to obtain resistance to the grand coalition, by definition, it suffices that the solution is Pareto-efficient (see Definition~\ref{def:pareto-efficient}).

\begin{Definition}[\bf Pareto-efficient Element]
\label{def:pareto-efficient}
Let $m\in \mathbb{N}$ and $S \subseteq \mathbb{R}_+^m$ be some set. A vector $\uu \in S$ is \emph{Pareto efficient} in $S$ if there doesn't exist another vector $\uu' \in S$ such that $u_t' \ge u_t$ on each coordinate and there exist $1\le t \le m$ s.t. $u_t' > u_t$. We denote the subset of Pareto efficient elements by $S^{PE}$.
\end{Definition}

\begin{Lemma}
\label{lemma:core-step2}
Let $\calU$ be the set of possible utility vectors for defenders in $\calN$. That is, $\uu \in \calU$ if there exist a coverage vector $\cc \in \calC_{k_\calN}$ such that $\uu = (U_i^d(\cc))_{i \in \calN}$. Let $\uu \in \calU^{PE}$, and let $\cc \in \calC_\calN$ be a coverage vector inducing $\uu$. Then $\CS=\langle \calN,\cc \rangle$ is a coalition structure resistant to any successful deviation of the grand coalition $D=\calN$.
\end{Lemma}
\begin{proof}
Assume in contradiction that there exist a successful deviation of $(\calN,\xx_D)$. Denote by $\uu'=(U_i^d(\xx_D))_{i \in \calN }$. Then by definition of a successful deviation, for every deviator $i \in \calN$, $u_i' > u_i$, contradicting the Pareto efficiency of $\uu$, since $\uu' \in \calU_{\calN}$ as well.
\end{proof}

\subsection{Inducing Arbitrary Attack Distributions}
\label{sec:proof-abf-surjectivity}
Reasonably, a coalition structure in the core should still result with a coverage vector $\cc$ with $\hgt({\cc}) \approx \overline{u}$, by the same logic that strategy profiles in NE should be efficient. Namely, if $\hgt(\cc) \gg \overline{u}$, the defenders may increase the height of the coverage and improve their utility, since all of the targets are more protected.

However, due to the presence of our ABF, allegedly it may be the case that some attack distribution $\omega(\cc)=p$ that can be generated at $\hgt(\cc)=u$, cannot be approximated by any coverage vector of height $\overline{u} \le \hgt(\cc) \le \overline{u}+\calO(\delta)$. We claim that this is never the case. Formally, we prove Proposition~\ref{prop:abf-surjectivity}.

\begin{proof}
For any subset $S\subseteq\calT$, we denote by $\omega_S(\uu)=\sum_{t\in S} \omega_t(\uu)$ the probability that the attack will take place in $S$. For every $t\in S$, we denote by $\restr{\omega}{S,t}(\uu):=\omega_t(\uu)/\omega_S(\uu)$ the probability that the attacker will attack target $t$ given that the attack will take place in $S$. We denote for every $m'\in\mathbb{N}$ the set $[m']:=\{1,\dots,m'\}$. We also denote by $p_S:=\sum_{t\in S} p_t$ and $p|_{S,t}:=p_t/p_S$.

W.L.O.G assume that $p_1 \le \dots \le p_m$. We prove by induction on $m$ the following: for every $1\le m' \le m$, there exists a utility vector $\uu$ such that:
\begin{enumerate}
    \item $\|\omega|_{[m'],t}(\uu) - p(t|[m'])\|_\infty \le 0.5m'(m'-1)\epsilon$
    \item $\forall t \in [m']: u \le u_t < u+2m'\delta$
    \item $|\omega|_{[m'],m'} - p|_{[m'],m'}| \le (m'-1)\epsilon$
    \item $\forall t \in [m']: |u_t-(u+m\delta)|<m'\delta$
\end{enumerate}

If $m'=1$ the claim is trivial taking $\uu=(u+m'\delta,0,\dots,0)$. Assume the claim is true for $m'-1$, and let $\uu$ be such a utility vector.

By Axiom~\ref{axiom:abf-independence}, if we extend $\uu'$ to a utility vector $\uu'=(u_1,\dots,u_{m'-1},u_{m'},0,\dots,0)$, we still have that $\omega|_{[m'-1],t}(\uu')=\omega|_{[m'-1],t}(\uu)$ for every target $t\in [m'-1]$.

If we take $u_{m'}=u_{m'-1}+\delta$ then $\omega|_{[m'],m'}(\uu') > 1-(m'-1)\epsilon$, as all the rest of the targets are attacked w.p. $<\epsilon$.

If $u_{m'}=u_{m'-1}-\delta$, then by Definition~\ref{axiom:abf-sensitivity} $\omega|_{[m'],m'}(\uu')<\epsilon$.

Therefore, by continuity, there exists a value $u_{m'}$ satisfying $|u_{m'}-u_{m'-1}|<\delta$ such that $\omega|_{[m'],m'}(\uu')$ is any value in the range $[\epsilon,1-(m'-1)\epsilon]$. Therefore, there exists such a value satisfying $|\omega|_{[m'],m'}(\uu') - p|_{[m'],m'}|<(m'-1)\epsilon$, as desired. Since by induction hypothesis $ |u_{m'-1} - (u + m\delta) | < (m'-1)\delta$, by triangular inequality we have that $|u_{m'}-(u+m\delta)|<m'\delta$ as desired. Since we didn't change the values of the rest of the targets in $[m']$, the only part left to check is that they also have the desired probabilities. Let $t<m'$ be some target. Then:

\begin{equation}
\begin{aligned}
&|\omega|_{[m'],t}(\uu')-p|_{[m'],t}| = |\frac{\omega|_{[m'-1],t}(\uu')}{\omega|_{[m'-1]}(\uu')} \omega|_{[m'-1]}(\uu')\\
&-p|_{[m'],t}| \\
&\le |\frac{\omega|_{[m'-1],t}(\uu)}{\omega|_{[m'-1]}(\uu)} - p|_{[m'-1],t}|\cdot \omega|_{[m'-1]}(\uu) \\
&+ |\omega|_{[m'-1]}(\uu) - p|_{[m'-1]}|\cdot p|_{[m'-1],t} \\
&\le 0.5(m'-1)(m'-2)\epsilon + (m'-1)\epsilon \\
&= 0.5m'(m'-1)\epsilon
\end{aligned}
\end{equation}

Therefore, we get that $\| \omega(\uu')- p(\cdot|t\le m') \|_\infty \le 0.5m'(m'-1)\epsilon$. Therefore, by induction, the claim is also true for $m'=m$, which completes the proof.

\end{proof}

This proves that if $\overline{\cov}(v)$ is feasible, then by decreasing the coverage of each target by $\calO(\delta)$, it is possible to $\calO(\epsilon)$ approximate any attack distribution $p$. This will result with an $\calO(\delta+\epsilon)$ utility loss for each defender. In what follows we combine all of our observations to construct stable solutions.




\subsection{A Construction of an \texorpdfstring{$\mathbold{\alpha}$ $\mathbold{\zeta}$}{}-Core Solution}
By this stage, we have everything needed to construct a solution to the $\alpha$ $\zeta$-core. We first introduce an approximation-preserving reduction into a simpler problem which can be solved accurately. To do so we again overload notations and define the utility when the attack probability is specified:

\begin{Definition}[\bf Defender Utility for Specified Attack Distribution]
Let $p$ be some probability distribution over $m$ targets, let $i \in \calN$ be some defender and $\cc \in \calC_{k_i}$. Then $U_i^d(\cc,p):=\sum_{t\in \calT}{p(t)\cdot U_i^d(\cc,t)}$.
\end{Definition}

\begin{Lemma}
\label{lemma:eps-p-equivalence}
Fix $\CS_{GC}$ to be the grand coalition structure with min-max height. There exist $A,B,\delta_0,\epsilon_0>0$ s.t. for every $\delta < \delta_0$, $\epsilon < \epsilon_0$, the following is true:

\begin{itemize}
    \item $\forall$ probability distribution $p$, $\exists$ coalition structure $\CS$ s.t. $\forall i \in \calN: U_i^d(\CS_{GC}, p) \le U_i^d(\CS) + A\epsilon + B\delta$.
    \item $\forall$ coalition structure $\CS$, $\exists$ probability distribution $p$ s.t. $\forall i\in\calN: U_i^d(\CS) \le U_i^d(\CS_{GC}, p) + A\epsilon + B\delta$.
\end{itemize}

\end{Lemma}

\begin{proof}
Note that the proof is constructive, and the reductions are efficient. Basically, the first direction follows from Proposition~\ref{prop:abf-surjectivity}, and it is efficient due to Axiom~\ref{axiom:abf-monotonicity}. As for the second direction, take $p:=\omega(\cov(\CS))$. Any target $t \in \calT$ with $U^a(\CS,t) < \overline{u}-\delta$ is attacked with probability less than $\epsilon$, therefore it would contribute a utility loss bounded by $r_i^d(t)\epsilon$. For the rest of the targets, since $\hgt(\CS_{GC})=\overline{u}$, the utility loss is bounded by $\delta$.
\end{proof}



\begin{Theorem}
\label{theorem:nonemptyness-eps-core}
There exists a game dependent $A,B,\delta_0,\epsilon_0>0$ such that for every $0<\delta<\delta_0,0<\epsilon < \epsilon_0$ and every $(\delta,\epsilon)$-ABF, the $\alpha$ $\zeta=(A\epsilon+B\delta)$-core is non-empty, and there exists an efficient algorithm that finds stable solutions in it.
\end{Theorem}

The main idea is the following: we start with a $\CS$ resistant to any deviation set $D\subseteq \calN$ with $k_D<k_\calN$ resources. Denote by $\uu^0:=(U_i^d(\CS))_{i \in \calN_+}$. We then search for a Pareto efficient solution in $\calU_{Pr}:=\{(U_i^d(\CS_{GC},p))_{i \in \calN_+} | p \text{ is a probability distribution}\}$ which satisfies the constraints $U_i^d(\CS_{GC},p) \ge \uu_i^0$ for each defender $i \in \calN_+$. We know such a solution $\CS'$ exists, since $p = \mathbb{1}_{t^*}$ satisfies the constraints.
There are three key steps left to finish the proof:
\begin{itemize}
    \item The solution $\CS'$ is resistant to any $\epsilon'$ successful deviation of any $D\subseteq \calN$ defenders with $k_D < k_\calN$ resources. $\epsilon'$ is the same as in Theorem~\ref{theorem:core-step1}. This follows directly from the fact that $\CS'$ realizes the constraints and because $\CS$ is resistant to such deviations.
    \item The solution $\CS'$ is resistant to any deviation set $D\subseteq \calN$ of $k_D = k_\calN$ resources. This follows directly from the Pareto efficiency of the solution.
    \item The solution $\CS'$ can be found efficiently. In order to see that, denote for each $t\in \calT$ by $\uu^d_t:=(U_i^d(\CS_{GC},t)_{i \in \calN_+}$. Then $\calU_{Pr}$ is a polytope with vertices at $\{\uu^d_t | t \in \calT \}$ (or a subset of that). Furthermore, the constraints for the defenders are linear with the variables $p(t)$ which span this polytope. Therefore, if we take any vector $\boldsymbol{\alpha}:=(\alpha_i)_{i \in \calN_+}$ with all entries $>0$, then maximizing $\langle \boldsymbol{\alpha}, \uu \rangle$ for $\uu \in \calU_{Pr}$ under the specified constraints is an LP optimization problem with $\calO(m+n)$ constraints and $m$ variables, therefore it can be solved efficiently, and yield the resulting solution.
\end{itemize}

The last step is to perform our approximation reduction step using Lemma~\ref{lemma:eps-p-equivalence}. In this step we gain an additional $\calO(\delta+\epsilon)$ utility loss, and the constants add-up.

\section{Examples}

\subsection{Example of an ABF}
\label{sec:ABFexample}
As mentioned, a well studied private case for an attacker ABF is the \emph{soft-max} function:
$$ \text{soft-max}_t(\uu):=\frac{e^{u_t}}{\sum_{t'\in\calT}{e^{u_{t'}}}} $$
In the domain of security games it is often called \emph{quantal response}, as the more utility the attacker gets from a target, the more the likelihood of being attacked increases, respectively.

\begin{Proposition}
The quantal response function $\omega$ is an ABF.
\end{Proposition}

\begin{proof}
~\paragraph{Probability Distribution} Let $\uu$ be some attacker utility vector and let $t\in\calT$. Denoting by $A = \sum_{t\in\calT}{e^{u_t}}$:
$$ \sum_{t\in\calT} {\omega_t(\uu)} = \sum_{t\in\calT}{e^{u_t}}/A = A / A = 1 $$
and also $0 \le \omega_t(\uu) \le 1$ since $exp(x)\ge 0$ for any $x \in \mathbb{R}$.

~\paragraph{Monotonous} Let $\delta>0$, $t\in\calT$ and define $\uu'$ such that $u_t'=u_t+\delta$ and $u_{t'}=u_{t'}$ for any other $t'\in\calT$. Denote by $A' = \sum_{t\in\calT}{e^{u'_t}}$, then $A' > A$. Also denote by $B = \sum_{t\neq t'\in \calT}{e^{u_t}} = \sum_{t\neq t'\in \calT}{e^{u'_t}}$. Then:
$$ \omega_t(\uu) = 1-(1-\omega_t(\uu)) = 1 - B/A < 1-B/A' = \omega_t(\uu') $$
~\paragraph{Independence} Let $\emptyset \neq S \subseteq \calT$ and let $t \in S$. Let $\uu'$ be some attacker utility vector such that for each $t'\not\in S$ $u'_{t'}=u_{t'}$. Then, denoting by $C=\sum_{t\in S}{e^{u_t}}=\sum_{t\in S}{e^{u'_t}}$:
$$ \frac{\omega_t(\uu)}{\sum_{t'\in S}{\omega_{t'}(\uu)}} = \frac{e^{u_t}/A}{C/A} = \frac{e^{u_{t'}}/A'}{C/A'} = \frac{\omega_t(\uu')}{\sum_{t'\in S}{\omega_{t'}(\uu')}}$$
\end{proof}
~\paragraph{Continuously Differentiable} Holds as a summation, division and composition of continuously differentiable functions.

Note that this would have also worked for any $\omega_t(\uu)=\frac{f(u_t)}{\sum_{t'\in\calT}{f(u_{t'})}}$ where $f$ is a one-dimensional monotonic continuously differentiable function into $(0,\infty)$.

Next we will construct a $\delta, \epsilon$ ABF.

\begin{Proposition}
\label{prop:quantal-response-abf}
Let $\delta,\epsilon > 0$. Then there exists a $(\delta,\epsilon)$-ABF.
\end{Proposition}

\begin{proof}
Consider $\omega(\uu) := \text{soft-max}(n\uu)$ for some $n\in \mathbb{N}$ to be determined later. By Proposition~\ref{prop:quantal-response-abf} $\omega$ is an ABF. Consider two targets $t,t'$ where $u_{t'} < u_t - \delta$, then:

$$ \omega_{t'}(\uu) < \frac{e^{n(u_t-\delta)}}{\sum_{s\in\calT}{e^{n u_s}}} < e^{-n\delta} $$

Therefore, we want $e^{-n\delta} < \epsilon$, and so $n>\frac{1}{\delta} \ln{\frac{1}{\epsilon}}$. 
\end{proof}

\subsection{Example of an SSG with no NSE}

\begin{Example}[\bf NSE may not Exist]
\label{example:no-wse}
Consider a single defender SSG with two targets. The attacker is indifferent to the targets. Assume for instance that the reward and the penalty of the attacker on both targets are $1,0$ respectively. The defender has two security resources, and prefers target $t_1$ over $t_2$. Assume for instance that $r^d(t_1) = 3 > p^d(t_1) = 2>r^d(t_2) = 1 > p^d(t_2) = 0$. Then, assuming the attacker breaks ties in opposition to the defender, or even attacks randomly over $\BR$, there is no NE for this game.
\end{Example}

\begin{proof}
First, observe that the defender can get a utility arbitrarily close to $r^d(t_1)=3$. Indeed, let $0 < \epsilon < 1$ and consider the coverage vector $(1-\epsilon,1)$, which is feasible for the defender. Then $\BR = \{t_1\}$, therefore the utility of the defender will be $3-\epsilon$.

We claim that there is no strategy for the defender to gain a utility $\ge 3$. If that is the case, then there is no NE, since the defender can always improve his utility to be a little closer to $3$.

Indeed, assume in contradiction that the defender gets a utility  $\ge 3$ playing some strategy $\cc$. Let $p$ be the probability that the attacker takes target $t_1$. Then $3 \le U^d(\cc)=p\cdot U^d(\cc,t_1) + (1-p)\cdot U^d(\cc,t_2) \le p\cdot r^d(t_1) + (1-p)\cdot r^d(t_2) = 2p + 1 \le 3 $. Therefore, $p$ must be equal to $1$ and so $c_1 < c_2 \le 1$. However, in that case, we get that $3=U^d(\cc,t_1)=c_1 \cdot r^d(t_1) + (1-c_1) \cdot p^d(t_1) = 2+c_1 < 3$, which is a contradiction.

\end{proof}

Note that the fact that the defender can get a utility arbitrarily close to $3$, but cannot get exactly $3$, is due to the fact that the utility is not continuous. Specifically, we took advantage of the fact that:
$$ 3 = \lim_{c_1 \to 1^{-}} U^d((c_1,1)) \neq U^d((1,1)) = 1 $$

In the case that the attacker plays randomly over equally appealing targets in $\BR$, we would still get $U^d((1,1))=0.5\cdot 3 + 0.5\cdot 1 = 2 \neq 1$, and a similar argument works.

\subsection{Example of a Robust Solution}
Consider the game described in Example~\ref{example:no-wse}. For single defender games, due to the phenomena described in that example, it is well accepted that the attacker breaks ties in favor of the defender. A NE under this assumption is called SSE; it is known to exist and can efficiently be found using LP. However, as we will now demonstrate, it is extremely sensitive and is not robust.

We begin by demonstrating how exactly this optimistic tie-breaking fails to form robust solutions even for single defender games, and then show how our model for SSG handles this problem.

\begin{Proposition}
Consider the game described in Example~\ref{example:no-wse}, assuming that the attacker breaks ties in favor of the defender. Let $0<\delta<1$. Then any NE of this game is not $\delta$-robust.
\end{Proposition}

\begin{proof}
We have already shown that the utility of the defender is bounded by $3$, and now that we assume optimistic tie breaking there is also a unique strategy that achieves that, namely $\cc_{NE}=(1,1)$, which is therefore the only NE. Then if we assume, for instance, that the parameters of the attacker on target $t_2$ slightly change to be $r^a(t_2)=\delta/2,p^a(t_2)=\delta/2$ and the parameters of target $t_1$ stay the same, then in the resulting game the defender gets a utility of $\delta/2$, and therefore it's not NE as the defender can deviate and get a utility of $3$.
\end{proof}

This formally explains the more intuitive argument, that when a solution assumes optimistic tie-breaking, a slight change in the attacker's parameters can lead to major utility loss for the defenders, altering the NE. This happens due to the non-continuity of the defender's utility at the equilibrium point. However, when the attacker follows an ABF and when the two targets are covered equally, they are attacked with some probabilities $p,q$ (50\%-50\% in quantal response for example), and slightly changing the attacker's parameters ($\delta$ parameter) will only slightly change the attacker's probability of attacking each target ($\epsilon$ parameter). This is the property that leads to robustness.

Next we are going to demonstrate a robust approximate NE of the above game following our constructive proof for Theorem~\ref{theorem:eps-NE-construction}. To make the game canonical for the sake of simplicity, assume that the defender has a single resource. Therefore, $\overline{u}=0.5\in(0,1]=(\overline{p},\overline{r})$.

Following the constants of the proof, we have $\delta_0=1/64$ (maximal value of $\delta$ such that the game is $\delta$-canonical), $\epsilon_0=1/2$. With respect to the parameters of the game, observe that these values are not that small, as solving with an accuracy of 2/3 of a decimal point for game parameters in the range of $[0,10]$ seems both common and feasible. Therefore, assume that the attacker's ABF for this game has parameters $\delta=0.01 < \delta_0$ and $\epsilon=0.1 < \epsilon_0$. For example, consider $\omega(\uu)=\text{soft-max}(n\uu)$ with $n=250$ (see Proposition~\ref{prop:quantal-response-abf}). However, the proof itself doesn't require knowing the exact ABF $\omega$, it is enough to know its parameters.

Therefore, we know from the proof that the solution would look something like $\cc_0=(0.5-\alpha, 0.5+\alpha)$ which, assuming $\alpha > \delta/2$, will induce the attacker to attack target $t_1$ with a probability of at least $1-\epsilon$. Therefore the utility of the defender is at least $(1-\epsilon)(2+(0.5-\alpha))=2.5-2.5\epsilon-\alpha+\alpha\epsilon > 2.5-2.5\epsilon-\alpha$, which we will denote by $U_0$.

Based on this theorem, we can bound the utility of a defender in this game by $U < \max{2 + (0.5+\delta/2), 3\epsilon + 1} = 2.5+\delta/2$.

For this example game it is intuitive: either $c_1 < c_2 - \delta$, in which case $c_1$ is limited to $0.5+\delta/2$, or the opposite is true, which would mean that target $t_1$ is attacked with a probability of at most $\epsilon$. We can bound the $\omega_{t_1}(\cc)$ by $1$ in the first case, and $U_1^d(\cc,t_1)$ by $r_1^d(t_1)$ for the second case.

Therefore, taking for instance $\alpha=\delta$ as the theorem suggests, $\cc_0$ is a $\zeta=2.5\epsilon+1.5\delta=0.265$-NE. Thus, following Theorem~\ref{theorem:delta-robustness}, $\cc_0$ is a $\delta=0.01$-robust $\zeta'=0.665$-NE.

Also note that as it is a single defender game, $\cc_0$ is also a $\delta$-Robust $\zeta'$ solution to the $\alpha$-Core (and also $\beta,\gamma$-Cores which coincide for a single defender).

\begin{Example}[\bf Grand Coalition Deviation of Type \ref{dev:induce-covered-targets}]
\label{example:gc-deviation-induce-covered-targets}
Consider the SSG depicted in \ref{table:not-stable}. Let us now use procedure \textsc{GC-ALLOC} with input $(\overline{u}-A\delta,\overline{u})$. We claim that the grand coalition has a deviation of type \ref{dev:induce-covered-targets} from the resulted strategy profile.
\end{Example}

Defender $1$ will first cover target $t_1$, which is the less favorable target for him, up to height $u$. He will then cover part of target $t_3$. Afterwards, defender $2$ will cover target $t_2$, which is less favorable for him, and then he will use the rest of his resources to cover target $t_3$, which will result with an attacker height $\overline{u}+\delta'$.

We take $A\delta>\delta$ so that deviation of type $\ref{dev:induce-uncovered-targets}$ will yield up to $\calO(\epsilon)$ utility gain for the defenders, but small enough such that $\ref{dev:increase-coverage}$ will yield up to $\calO(\delta)$ utility gain for the defenders. For example, we can take $A\delta=2\delta$.

\begin{table}[ht!]
\centering
\begin{tabular}{ |p{2cm}|p{1cm}|p{1cm}|p{1cm}||p{1cm}|  }
\hline
Player      & $t_1$   & $t_2$   & $t_3$ & $K_i$ \\
\hline\hline
Defender 1  & 1,1     & 300,300 & 2,2   & 1     \\
Defender 2  & 300,300 & 1,1     & 2,2   & 1     \\
\hline
Attacker    & 1,0     & 1,0     & 1,0   &      \\
\hline
\end{tabular}
\caption{Example of an SSG where using \textsc{GC-ALLOC} doesn't yeild a stable solution due to a deviation of type \ref{dev:induce-covered-targets}}
\label{table:not-stable}
\end{table}

Unfortunately, the grand coalition still has a deviation of type \ref{dev:induce-covered-targets} with $\Omega(1)$ utility gain for both defenders. In the original $\CS$, both defenders get a utility close to $2$. Consider a deviation of the grand coalition, with a coverage of $(0,0,1)$. The utility that each defender gets now is at least $0.5*300=150\gg2$. Therefore, this is a successful deviation.

Intuitively, the reason the method for the original game for finding the core doesn't work is that in this case defender $1$ covers a target that defender $2$ prefers over the chosen target, and vice versa. Together, they can induce the attacker to attack these preferred targets, each with probability 0.5. This wasn't possible in the NLE for the non-cooperative game by definition. Also, in the original model, every defender assumed that the attacker attacks the target that is worst for him, so this wouldn't be a successful deviation there either.

Since the reward and penalties on each target are the same in this game, it is convenient to specify which solutions are in the core by the distribution of the attacker. We claim that the distribution $(p,1-p,0)$ is in the $\epsilon$-core for any $\frac{2}{300}\le p\le 1-\frac{2}{300}$. Indeed, in such a strategy the defenders get utilities $300p, 300(1-p) \ge 2$ respectively. A deviation by a single defender $i$ is not successful, since the other defender can block the target $t_{3-i}$ and the utility of defender $i$ is bounded by $2$. A successful deviation of the grand coalition means that $u_1 > 300p, u_2 > 300(1-p)$. If the coverage vector of this deviation is $(q_1, q_2, q_3)$ then we must have $q_1 > p, q_2 > (1-p)$ and therefore $q_1+q_2+q_3>1$, which is impossible.

It seems as if the solutions we find are a subset of the Pareto efficient defender utility vectors, specifically the Pareto efficient solutions where no subset of defenders can deviate. This observation will turn out to be important for constructing an element in the core.

\subsection{An example of an SSG with an Empty Robust \texorpdfstring{$\gamma$}{}-Core}

In $\gamma$-core (see Definition~\ref{def:gamma-core}), we assume that, after deviation, the rest of the defenders are singletons and cannot cooperate, as suggested in~\cite{chander2010cores}. There are usually further limitations on the revenge takers' response, such as that each revenge taker best responds to a deviation~\cite{chander2006cooperation} or that the resulting strategy profile forms a NE~\cite{chander2007gamma}. However, without any of these assumptions, we show that the $\gamma$-core of the game may be empty.

\begin{Definition}[\bf Robust $\mathbold{\gamma}$-Core]
\label{def:gamma-core}
Let $\CS$ be some coalition structure. A deviation $(D,\xx_D)$ from $\CS$ is $\gamma$-successful if for every non-cooperative strategy profile of the rest of the defenders $Y=(\yy_r)_{r\in R}$, the following holds for every deviator $i \in D$:
$$ U_i^d(\langle \xx_D, Y \rangle) > U_i^d(\CS) $$

$\CS$ is in the $\gamma$-core if it has no $\gamma$ successful deviations.
\end{Definition}

We demonstrate that the $\gamma$-core may be empty.

\begin{Example}[Empty $\gamma$-Core]
\label{example:empty-gamma-core}
The SSG depicted in Table~\ref{table:empty-gamma-core} has an empty $\gamma$-core, assuming $(\frac{7-3\sqrt{5}}{12},\frac{1}{15})$-ABF.




\end{Example}

Basically, defenders $1,2$ can always $\gamma$-deviate and get a utility $\ge 4$ each, as can defenders $3,4$, but no coalition structure can yield all defenders $\ge 4$ utility. Here is a more formal and detailed explanation:

\begin{proof}
If defenders $1,2$ decide to play $(2/3,2/3,2/3,0,0,0)$ which can be implemented as $K_{\{1,2\}}=2$, then no matter what defenders $3,4$ decide to do, a target $t\in{t_1,t_2,t_3}$ is attacked w.p. at most $3\epsilon$. Indeed, if defenders $3,4$ want to incentivize the attacker to attack targets in ${t_1,t_2,t_3}$, they need to cover targets $t_4,t_5,t_6$ and result with a coverage with a diameter smaller than $\delta$. The best they can do non-cooperatively with their $K_{3,4}=2$ resources is to play $(0,0,0,\alpha,\alpha,\alpha)$ where $\alpha=1-\alpha^2$. This leads to an attacker height of $1-\alpha = \frac{3-\sqrt{5}}{2}$. However, $\delta < \frac{3-\sqrt{5}}{2} - \frac{1}{3}$.

Therefore, after deviation, defenders $1,2$ will get at least $(1-3\epsilon)\cdot p_1^d(t_4) = 5\cdot(1-3\epsilon)=4$.

On the other hand, by symmetry, it is easy to see that defenders $3,4$ can get at least a utility of $4$ after deviation as well, assuming defenders $1,2$ will respond without coordination. Therefore, if a coalition structure $\CS$ was in the $\gamma$-core, then it must yield a utility $U_1^d(\CS)=U_2^d(\CS)\ge 4$, since otherwise, we proved that there exists a deviation of defenders $1,2$. Note that since defenders $1,2$ are of the same type, their utility is always equal. For the same reason, $U_3^d(\CS)=U_4^d(\CS) \ge 4$. Denote by $p$ the probability that the attacker attacks a target in $\{t_1,t_2,t_3\}$. On the one hand, $4 \le U_1^d(\CS) \le p\cdot 1 + (1-p)\cdot 6$, therefore $p \ge 0.6$. On the other hand, $4 \le U_3^d(\CS) \le p\cdot 6 + (1-p)\cdot 1$, therefore $p \le 0.4$, a contradiction. Thus, there always exists a $\gamma$-successful deviation, and the $\gamma$-core is empty.
\end{proof}

It is worth mentioning that the $\gamma$ core for the original MSSG model is also empty, by a similar example.

\end{document}